\documentclass[A4paper,12pt]{article}
\usepackage[margin=2.2cm,nohead]{geometry}
\usepackage{graphicx}
\usepackage{amsfonts}
\usepackage{amssymb}
\usepackage{amsmath}
\usepackage[symbol]{footmisc}
\usepackage[english]{babel}
\usepackage{appendix}
\usepackage{lscape}
\usepackage{subcaption}
\usepackage{caption}
\usepackage{hyperref}
\usepackage{natbib}
\usepackage{setspace}
\usepackage{color} 
\usepackage{tabularx}
\newcommand{\ed}{\end{document}}
 
\begin{document}

\title{The Gender Pay Gap Revisited with Big Data:\\ Do Methodological Choices Matter?\thanks{We acknowledge helpful comments by Philipp Bach, Marina Bonaccolto-T\"{o}pfer, Christina Felfe, Martin Huber, Pat Kline, Michael Knaus, Matthias Krapf, and Michael Lechner, seminar participants at the University of Basel, University of Linz, and LISER, Luxemburg, as well as conference participants at EALE/SOLE 2020, Verein f\"ur Socialpolitik 2020, and the IAB 2020 workshop on ``Machine Learning in Labor, Education, and Health Economics''. Anthony Strittmatter gratefully acknowledges financial support from the Swiss National Science Foundation (Spark Project 190422) and the French National Research Agency (LabEx Ecodec/ANR-11-LABX-0047). The authors are solely responsible for the analysis and the interpretation thereof.}}

\author{Anthony Strittmatter\footnote{CREST-ENSAE, Institut Polytechnique Paris, France; CESifo, Munich, Germany; email: anthony.strittmatter@ensae.fr.} \and Conny Wunsch\footnote{Faculty of Business and Economics, University of Basel; University St.\ Gallen, Switzerland; email: conny.wunsch@unibas.ch.}}

\date{\today}

\maketitle
\thispagestyle{empty}

\begin{abstract}
\noindent 
The vast majority of existing studies that estimate the average unexplained gender pay gap use unnecessarily restrictive linear versions of the Blinder-Oaxaca decomposition. Using a notably rich and large data set of 1.7 million employees in Switzerland, we investigate how the methodological improvements made possible by such big data affect estimates of the unexplained gender pay gap. We study the sensitivity of the estimates with regard to i) the availability of observationally comparable men and women, ii) model flexibility when controlling for wage determinants, and iii) the choice of different parametric and semi-parametric estimators, including variants that make use of machine learning methods. We find that these three factors matter greatly. Blinder-Oaxaca estimates of the unexplained gender pay gap decline by up to 39\% when we enforce comparability between men and women and use a more flexible specification of the wage equation. Semi-parametric matching yields estimates that when compared with the Blinder-Oaxaca estimates, are up to 50\% smaller and also less sensitive to the way wage determinants are included. 

\medskip

 \medskip

\end{abstract}

%%%%%%%%%%%%%%%%%%%%%%%%%%%%%%%%%%%%%%%%%%%%%%%%%%%%%%%%%%%%%%%
\newpage
\setcounter{page}{1} \renewcommand*{\thefootnote}{\arabic{footnote}}
\section{Introduction}\label{sec:intro}

Achieving gender equality, especially equality in pay, is among the top priorities for governments in many countries. Measuring unexplained inequalities in pay between women and men, the so-called unexplained gender pay gap, has been the subject of an extensive literature for more than half a century \citep[see][for comprehensive reviews]{blau00,blau17,gold17b,kunz18,oli16}.
%\citep[see the reviews of, e.g.,][]{ww05,vts15}. 
A large literature focuses on understanding the sources of inequality in pay by studying the sensitivity of the gender pay gap with regard to the inclusion of important wage determinants.
%, including gender segregation by occupation, industry, and establishment \citep[e.g.,][]{blau00,bayard03,bar10,gold17,sin20}, child penalty \citep[e.g.,][]{ejr13,ang16,gold17b,bue18,kle19b,kle19,kra20}, and part-time penalty \citep[e.g.,][]{man08,gold14,liu16}.
A much smaller literature focuses on the impact of methodological choices. We contribute to the latter literature by investigating the sensitivity of the estimated gender pay gap with regard to three dimensions: enforcement of overlap in wage determinants across gender, flexibility of model specification, and estimation method. Our results suggest that these methodological choices can reduce the estimated unexplained gender pay gap by as much as 50\%, even when keeping relevant wage determinants fixed across estimates. This suggests that methodological choices are at least as relevant for gender pay gap estimations as considerations about wage determinants.

\citet{ww05} and \citet{vts15} document that the vast majority of existing studies use a linear version of the Blinder-Oaxaca decomposition \citep[BO,][]{BLI,OX} to estimate the mean unexplained gender pay gap \citep[see][for a comprehensive review of BO and other decomposition methods]{fort11}. Thus, BO estimates serve as the key input for policy makers who aim to achieve equality in pay. However, most applications of BO impose a number of restrictions that may not be realistic. Firstly, they typically use relatively inflexible functional forms for the wage equation. For example, the returns to education are often assumed to be the same across occupations, age and experience, which contradicts both theory and empirical evidence \citep[see, e.g.,][]{lem14}. Secondly, standard applications of BO do not account for common support violations, i.e., the lack of observationally comparable men for every woman. BO extrapolates into regions without support based on the assumed functional form. If the functional form is misspecified and there is lack of support, then this extrapolation may lead to bias. Thirdly, BO restricts heterogeneity in gender pay gaps to variable-specific differences in coefficients. Any heterogeneity that is not captured by this may bias estimates of the mean gender pay gap. This is particularly relevant, because heterogeneity in gender pay gaps is widely acknowledged in the literature \citep[see, e.g.,][]{bach18,bar10,bon01,cfm13,chern18,goldin14}. 

Of course, these restrictions can easily be relaxed when working with large data sets. Researchers can model the inclusion of control variables in more flexible ways. They can check and enforce common support ex ante, even for parametric estimators like BO. Furthermore, they can choose more flexible estimators that unlike BO do not restrict pay gap heterogeneity. Surprisingly though, these adjustments are rarely implemented in applied work. Therefore, an important question we explore in this study is whether these adjustments really matter in practice. To answer this question, we exploit a very large data set that covers about 37,000 establishments with individual data on more than 1.7 million employees in Switzerland, which covers almost one third of all Swiss employees. The data allow us to take full advantage of the methodological improvements that are possible with existing methods. In particular, they allow us to go as far as exact matching on all elements of the rich set of observed wage determinants with resulting cells that are still large enough for meaningful analysis after enforcing full support. 

We start by analyzing common support and the gender pay gap obtained from exact matching applying the technique of \citet{nopo08}. Thereafter, we estimate the mean unexplained gender pay gap with different parametric and semi-parametric methods and for various samples that differ in how strictly we impose common support. As estimators, we consider a linear regression model (LRM) with a dummy for women, BO, inverse probability weighting (IPW), augmented IPW (AIPW) as a doubly robust mixture between BO and IPW, propensity score matching (PSM) and a combination of exact matching and PSM (EXPSM). For each estimator, we consider three model specifications that differ in how flexibly we include the observed wage determinants: (i) the \textit{baseline model} contains dummy variables for all categories and quadratic terms for the continuous variables (up to 117 control variables), (ii) the \textit{full model} additionally contains higher-order polynomials as well as a large number of interactions between wage determinants (up to 615 control variables), and (iii) the \textit{machine learning model} employs LASSO estimation techniques \citep{tib96} to select the relevant wage determinants in a data-driven way \citep[see][for an introduction to LASSO]{hast09}. Specifically, we implement the post-double-selection procedure \citep[][]{bell13}, double-machine-learning \citep{che17}, and T-learner \citep{kuenz19}. Furthermore, we study the private and public sectors separately. The two sectors are subject to different degrees of labor market regulation and they attract different types of workers, both which influence the size of observed gender pay gaps \citep[see, e.g.,][]{bar10}. 

With this study, we contribute to the existing literature in several important ways. Firstly, we conduct the first comprehensive analysis of common support in the context of the gender pay gap. Secondly, we are the first to study how functional form restrictions regarding the inclusion of wage determinants affect pay gap estimates both within and across different estimators. By including machine learning techniques for data-driven model specification for all estimators, we also employ methods that have never before been considered in the context of the gender pay gap. Thirdly, we study a much more comprehensive set of estimators than any previous study. Finally, we are the first to consider all of these dimensions collectively and to vary them systematically. In total, we estimate the unexplained gender pay gap for five definitions of common support, six estimators, three model specifications, and two sectors, resulting in a total of 180 different estimates. This allows us to isolate the impact of each dimension and to understand their interactions.

The literature that investigates the sensitivity of the gender pay gap with regard to common support and estimation methods is small. \cite{nopo08} studies support violations and the resulting unexplained wage gaps with respect to four exemplary wage determinants with data from Peru. Moreover, he compares exact matching (EXM) with various linear specifications of BO that differ in how flexibly three exemplary wage determinants are included, and whether or not common support is enforced with respect to these variables. \cite{black08} and \cite{gtv17} compare BO and EXM estimates of the gender pay gap. \citet{dr07}, \citet{f07}, and \cite{mpw20} compare BO estimates of the gender pay gap with different propensity score matching estimators. Relatedly, \cite{barsky02} and \cite{bryan16} study the sensitivity of black-white gaps with regard to different parametric and semi-parametric estimators. \cite{bach18} and \cite{brie20} apply machine learning methods for model specification using the post-double-selection procedure. However, none of these studies provides a comparable large-scale systematic analysis of gender pay gap estimates with regard to our considered methodological choices.  
 
Our paper is related to other strands of the literature that investigate the sensitivity of the gender pay gap with regard to {other} dimensions. One strand of literature focuses on the assumptions required for the identification of gender discrimination. For example, there are studies that investigate biases due to gender-specific selection into employment \citep[e.g.,][]{chan19,chern20,maa19,mach17,no04,oli08} and potential endogeneity of the observed wage determinants \citep[e.g.,][]{kunz08,huber15,hub20,y15}. We maintain the same identifying assumptions for all our estimates. Thus, differences in our estimates result from the empirical methods and not from the underlying assumptions. To be precise, we measure the unexplained gender pay gap as the expected relative wage difference of employed women in the sample with support, compared to employed men with the same observed wage determinants. This {parameter} is informative about equal pay for equal work taking individual choices as a given. It does not necessarily measure gender discrimination in pay because gender discrimination might influence eventual pay earlier in life through, for instance, educational and career choices. Moreover, there may be unobserved factors that help explain observed gender differences in pay. A large strand of literature identifies wage determinants that contribute to explaining the gender pay gap and analyzes how including these wage determinants changes estimates of the unexplained gender pay gap. We discuss this literature in Section \ref{sec:data2} when we describe the variables we observe in our data. In our study, we keep the wage determinants for which we control constant across all estimates and only vary how flexibly we include them in the estimations. 

We find that all the methodological choices we consider significantly impact the size of the estimated unexplained gender pay gap. The estimates decline by up to 50\% when stricter support is enforced, functional forms are relaxed, and less restrictive estimators are used. The lack of comparable men for each woman is a particularly serious issue. For 89\% and 70\% of employed women in the private and public sector, respectively, there is no support with regard to at least one observed wage determinant. For the public sector, we find that lack of support directly explains a large part of the raw gender pay gap. Moreover, with estimates that are 6-50\% lower, enforcing support strongly affects the estimates of the unexplained gaps for all estimators. Thus, checking support in applications ex ante and deciding on how strictly it should be enforced is crucial to applied work. 

With estimates that are around 5-19\% lower, the flexible inclusion of wage determinants is very important for the parametric estimators LRM and BO as well as the hybrid AIPW. In contrast, model specification has little impact on the results from semi-parametric matching estimators that do not model the wage equation. These estimators do not restrict heterogeneity in unexplained wage gaps, which we find to be important as well. Compared to the most flexible version of BO and with a reasonable choice of common support, flexible EXPSM reduces the estimated unexplained wage gap by another 14\% in the private sector and 35\% in the public sector. The results for semi-parametric IPW are ambiguous, which might be related to propensity score values close to one \citep[e.g.,][]{ka10,bus14}.

Based on our findings, we recommend enforcing common support with respect to the most important wage determinants ex ante. To estimate the unexplained pay gap, we recommend combining exact matching on {some important wage determinants} with radius matching on a flexibly specified propensity score. This minimizes the risk of functional form misspecification and offers a reasonable balance of comparability, precision of the estimate, and representativeness of the study sample. For the private and public sector, respectively, implementing these recommendations with our data explains 16\% and 43\% more of the raw wage gap than standard BO estimates and results in estimated unexplained pay gaps that are 23\% and 50\% lower. An important takeaway for policy makers is that the commonly reported BO estimates of the gender pay gap can be misleading.

The paper proceeds as follows. The next section describes the data we use. In Section 3, we explain our empirical model and estimation methods. Section 4 contains the results, starting with the analysis of common support, then followed by the results for the mean unexplained gender pay gap as a function of the three dimensions of methodological choices we consider. In Section 5, we discuss the generalizability of our results and provide recommendations for applied work. The last section concludes. Online appendices A-D contain supplementary material.

\section{Data}\label{sec:data}

\subsection{Sample selection}\label{sec:data1}

We use the 2016 wave of the Swiss Earnings Structure Survey (ESS). The ESS is a bi-annual 
survey of approximately 37,000 private and public establishments with individual data on more than 1.7 million employees, representing almost one third of all employees in Switzerland. The survey covers salaried jobs in the secondary and tertiary sectors in establishments
with at least three employees. Sampling is random within strata defined by establishment size, industry, and geographic location. Participation in the survey is compulsory for the establishments. The gross response rate is higher than 80\%. All results we report take into account the sampling weights provided by the ESS that correct for both stratification, and non-response. Typically, establishments report the required information directly from their remuneration systems. Thus, the survey effectively includes administrative data from establishments.

We restrict the analysis to the working population aged between 20 and 59 years (dropping 127,298 employees). We exclude employees for whom we observe very little support between men and women ex ante. This excludes 70,052 employees with less than 20 percent part-time employment, 2,706 members of the armed forces or agricultural and forestry occupations,\footnote{Based on 2-digit ISCO-08 codes 01, 02, 03, 61, and 62.} and 3,025 employees from the agricultural, forestry, mining, and tobacco sectors.\footnote{Based on 2-digit NOGA 2008 codes 01, 02, 05, and 12.} We analyze the gender pay gap separately for the private and public sector. The public sector offers more regulated wages and attracts a different selection of employees. For example, women are over-represented in the public at 56\%, but constitute only 43\% of the private sector. Moreover, the public sector is much more homogeneous in terms of industries and occupations. The baseline sample contains 1,132,042 employees in the private sector and 405,448 employees in the public sector, after dropping an additional 54 employees in industries with very few observations in the public sector.\footnote{Based on 2-digit NOGA 2008 codes 10, 31, 45, 47, 55, and 58.}

\subsection{Data description}\label{sec:data2}

The main variable of interest is a standardized wage measure that is provided by the Federal Statistical Office as part of the ESS. It measures the monthly full-time-equivalent gross wage including extra payments. The latter comprise add-ons for shift, Sunday, and night work, other non-standard working conditions, and irregular payments such as bonuses and Christmas or holiday salaries, but they exclude overtime premia. Wages are standardized to a 100\% full-time equivalent without overtime hours.

\begin{table}[htbp]
 \centering
  \caption{Means and standardized differences of selected variables} \label{bal}
	\begin{footnotesize}
    \begin{tabularx}{\textwidth}{Xccccccc} \hline \hline
          & \multicolumn{3}{c}{Private Sector} && \multicolumn{3}{c}{Public Sector} \\
       \cline{2-4} \cline{6-8}   & \multicolumn{2}{c}{Mean}   & Std.  && \multicolumn{2}{c}{Mean}  & Std. \\
          & Women & Men  & Diff. && Women & Men  & Diff. \\
  \cline{2-8}        & (1)   & (2)   & (3)   && (4)   & (5)   & (6) \\ \hline
  \multicolumn{8}{c}{Remuneration characteristics} \\\hline
            Standardized monthly wage (in CHF)    & 6,266 & 7,793 &     31.9    &       & 7,731 & 8,985 & 42.2 \\
             Irregular payments  &  .330   &  .411   &   16.9  &       & .138  & .227  & 23.3 \\\hline
    \multicolumn{8}{c}{Demographic characteristics} \\\hline
        Age    &         &         &         &       &         &         &    \\
       \quad 20-29 years  &  .206   &  .186   &   5.1   &       & .159  & .116  & 12.5 \\
       \quad 30-39 years  &  .270   &  .285   &   3.4   &       & .266  & .250  & 3.7 \\
       \quad 40-49 years  &  .275   &  .282   &   1.5   &       & .282  & .300  & 3.9 \\
       \quad 50-59 years  &  .249   &  .248   &   .4    &       & .293  & .335  & 8.9 \\
        Education  &         &         &         &         &       &         &     \\
       \quad Higher  &  .284   &  .320   &   7.9   &       & .539  & .570  & 6.2 \\
       \quad Vocational  &  .474   &  .466   &   1.7   &       & .279  & .288  & 2.1 \\
       \quad No vocational  &  .187   &  .168   &   4.9   &       & .074  & .048  & 11.0 \\\hline
    \multicolumn{8}{c}{Employment characteristics} \\\hline
              Tenure  &         &         &         &       &         &         &     \\
       \quad $<$ 2 years  &  .292   &  .262   &   6.8   &       & .191  & .161  & 8.0 \\
       \quad 2-4 years  &  .250   &  .233   &   4.0   &       & .218  & .197  & 5.2 \\
       \quad 5-7 years  &  .143   &  .138   &   1.6   &       & .193  & .178  & 3.9 \\
       \quad 8-15 years  &  .201   &  .209   &   2.0   &       & .234  & .262  & 6.6 \\
       \quad 16-45 years  &  .113   &  .158   &   13.2  &       & .164  & .202  & 9.8 \\
                  Management position  &         &         &         &       &         &         &     \\
       \quad Top  &  .031   &  .073   &   18.8  &       & .011  & .040  & 18.7 \\
       \quad Upper  &  .047   &  .076   &   12.0  &       & .075  & .109  & 11.6 \\
       \quad Middle  &  .079   &  .101   &   7.7   &       & .067  & .109  & 14.8 \\
       \quad Lower  &  .071   &  .083   &   4.3   &       & .057  & .077  & 8.2 \\
       \quad None  &  .771   &  .667   &   23.2  &       & .736  & .614  & 26.2 \\
            Share of women in occupation  &         &         &       &         &         &         &     \\
       \quad $<$ 25\%  &  .044   &  .297   &   71.4  &       & .028  & .107  & 32.0 \\
       \quad 25-50\%  &  .381   &  .518   &   27.7  &       & .296  & .440  & 30.3 \\
       \quad 50-75\%  &  .346   &  .158   &   44.3  &       & .395  & .278  & 24.9 \\
       \quad $>$ 75\%  &  .229   &  .027   &   63.2  &       & .214  & .131  & 22.2 \\
    Part-time and full-time work &       &       &       &       &       &       &  \\
   \quad 20-49\% & .195  & .038  & 50.6  &       & .187  & .059  & 39.8 \\
   \quad 50-79\% & .235  & .046  & 56.4  &       & .303  & .086  & 57.1 \\
   \quad 80-99\% & .164  & .066  & 31.0  &       & .221  & .124  & 25.9 \\
   \quad 100\% & .406  & .850  & 103.4 &       & .290  & .732  & 98.6 \\
        Establishment size  &         &         &         &       &        &         &     \\
       \quad $<$ 20 Employees  &  .255   &  .215   &   9.5   &       & .015  & .017  & 1.4 \\
       \quad 20-49 Employees  &  .131   &  .162   &   8.9   &       & .029  & .027  & 1.1 \\
       \quad 50-249 Employees  &  .237   &  .258   &   5.0   &       & .123  & .091  & 10.3 \\
       \quad 250-999 Employees  &  .142   &  .157   &   4.4   &       & .117  & .108  & 3.0 \\
       \quad $>$ 999 Employees  &  .236   &  .207   &   6.9   &       & .716  & .756  & 9.4 \\\hline
        Observations  & 491,007 & 641,035 &       &       & 227,617 & 177,831 &   \\\hline\hline
     \end{tabularx}
\parbox{\textwidth}{\footnotesize Notes: Table is based on the baseline sample before imposing sample restriction on support. The monthly regular wage is reported by employers. It is standardized to 100\% full-time equivalent wages without overtime hours. Table A.1 in the Online Appendix A provides a detailed description of all observed variables. \cite{ro83} classify absolute standardized difference (std. diff.) of more than 20 as ``large''.  }
\end{footnotesize}
\end{table}
    
The data contain a rich set of wage determinants that are considered important in the gender pay gap literature such as age, education, occupation, industry, establishment characteristics, wage bargaining, management position, marital status, and part-time work.\footnote{See \cite{bayard03}, \cite{bea14}, \cite{ber19}, \cite{ber10,ber01}, \cite{bre14}, \cite{brun19}, \cite{buf16}, \cite{card16}, \cite{fer14}, \cite{gob15}, \cite{gold17}, \cite{hei10}, \cite{liu16}, \cite{man08}, \cite{ober20}, \cite{sin20}, and \cite{win97} among others.} Table \ref{bal} documents the means and standardized differences of selected variables by sector and gender (see Table A.1 in Online Appendix A for a full list of all observed characteristics). In the private sector, the average standardized monthly wage is 6,266 CHF for women and 7,793 CHF for men.\footnote{The conversion rate of CHF to US dollars was approximately 1:1 in 2016.} In the public sector, the average wage is much higher at 7,731 CHF for women and 8,985 CHF for men. In both sectors, the share of women receiving irregular payments such as bonuses is notably smaller than that of men. This aligns with a similar difference in the share of employees holding a management position.

The distributions of age and education are relatively similar across gender in both sectors, although employed women tend to have slightly less education than men. Women are somewhat less likely to have long tenure and significantly less likely to work full-time than men. Labor market segregation by occupations is very strong. We illustrate this by showing the employment shares for occupations grouped by their shares of women. Unsurprisingly, more women work in female-dominated occupations than men and vice versa. Female-dominated occupations are more frequent in the public sector. Less than 3\% of men in the private sector, but more than 13\% of men in the public sector work in female-dominated occupations. Gender differences according to the establishments size show no strong systematic pattern.

While the data are quite rich with regard to available wage determinants, there are several important variables that we do not observe. One example is actual work experience \citep[e.g.,][]{coo20,gay12}. Potential experience is captured by age and education, and we also observe tenure. Hence, we capture some aspects of experience, but not all of it. Since experience is positively related to wages, and women have on average less work experience, we over-estimate potential violations of equal pay for equal work in Switzerland in this respect. However, other information that the literature emphasizes is missing as well. This includes, for example, competitiveness, children, environment during childhood, gender norms, and non-cognitive factors.\footnote{See, e.g., \citet{flo15} and \cite{gne09,gne03} for competitiveness,  \cite{add17}, \cite{and02}, \cite{ang16}, \cite{bai12}, \cite{bue18}, \cite{ejr13}, \cite{fitz16}, \cite{kle19b,kleven19}, \cite{kra20}, \cite{lund17}, and \cite{wald98} for children, \cite{aut19}, \cite{ber13}, and \cite{bren18} for environment during childhood, \citet{ber15} and \cite{rot18} for gender norms, and \citet{fort08} for non-cognitive factors.} Moreover, we account for neither selection into employment based on unobserved characteristics, nor for potential endogenity of control variables. Hence, our estimates are informative about equal pay for equal work when individual choices are taken as a given, subject to any omitted variable bias that may result from unobserved factors.

\section{Econometric model and estimators}
\subsection{Notation and parameter of interest}
We denote the gender dummy by $G_i$, with $G_i=1$ for employed women and $G_i=0$ for employed men. We use the logarithm of the standardized monthly wage as the outcome variable, which we denote by $Y_i$. The raw gender pay gap is
\begin{equation}\label{eq1}
\Delta = E[Y_i|G_i=1] - E[Y_i|G_i=0]. 
\end{equation}
The raw gender pay gap can be decomposed into an explained and unexplained part. The vector $X_i$ contains observed demographic and labor market characteristics of the employees as well as observed characteristics of their employers. The predicted wage of employed men, would they have the same observed characteristics as employed women is $E_{X|G=1}[ \mu_0(x) ]$, with $\mu_0(x) = E[Y_i|G_i=0,X_i=x]$. Adding and subtracting $E_{X|G=1}[ \mu_0(x) ]$ in (\ref{eq1}) gives
\begin{equation} \label{eq2x}
\Delta =   \underbrace{ E[Y_i|G_i=1] - E_{X|G=1}[  \mu_0(x) ]}_{\mbox{unexplained $\delta$}} +\underbrace{ E_{X|G=1} [  \mu_0(x) ]  - E[Y_i|G_i=0]}_{\mbox{explained $\eta$}}   . 
\end{equation}
The second difference on the right side of (\ref{eq2x}) is the part of the raw gender pay gap that can be explained by gender differences in the observed wage determinants $X_i$. The first difference on the right side of (\ref{eq2x}) is the gender pay gap for employed women that cannot be explained by gender differences in the observed wage determinants. It is the expected difference in pay of employed women compared to observationally identical employed men, which we denote by 
\begin{equation} \label{eq2}
\delta = E[Y_i|G_i=1] - E_{X|G=1}[ \mu_0(x) ].
\end{equation}
This is the parameter of interest in the majority of studies on gender wage inequality.

\subsection{Common support}

The unexplained pay gap (\ref{eq2}) is only identified if, for all females, there exist men that are observationally identical with respect to the wage determinants. Now assume that there is lack of support for some men or women. Let $S_{i}=1$ for individuals with support and $S_i=0$ for individuals without support. \citet{nopo08} shows that
\begin{equation*}\label{cs1}
\Delta =\underbrace{E[Y_i|G_i=1,S_i=1]-E[Y_i|G_i=0,S_i=1]}_{= \Delta_{S=1}}+\Delta^s_{G=1}-\Delta^s_{G=0}
\end{equation*}
where $\Delta^s_{G=g}\equiv Pr(S_i=0|G_i=g)[E[Y_i|G_i=g,S_i=0]-E[Y_i|G_i=g,S_i=1]]$ measures wage differences across individuals of gender $g$ (for $g\in\{0,1\}$) in and out of support. The parameter $\Delta_{S=1}$ is the raw wage gap within support, which we can decompose as
\begin{equation}\label{cs2}
\begin{array}{rl}
\Delta_{S=1} =&\underbrace{ E[Y_i|G_i=1,S_i=1] - E_{X|G=1,S=1}[  \mu_0(x)|S_i=1 ]}_{\mbox{$\delta_{S=1}$}}\\
& +\underbrace{ E_{X|G=1,S=1} [  \mu_0(x)|S_i=1 ]  - E[Y_i|G_i=0,S_i=1]}_{\mbox{$\eta_{S=1}$}}.
\end{array}
\end{equation}
The first right-hand term, $\delta_{S=1}$, is the unexplained gender pay gap for employed women with support. The second right-hand term, $\eta_{S=1}$, is the part of the raw gender pay gap with support that can be explained by gender differences in the wage determinants. 

We use the following procedure to analyze the influence of common support violations on gender pay gap estimates. First, we sequentially increase the number of wage determinants for which we impose common support and then study the share of females without support. Imposing common support based on a large number of wage determinants increases ex-ante comparability of women and men. Second, we estimate the elements of (\ref{cs2}) for each sequential step using an exact matching estimator (see Section \ref{match} for more details). Third, based on the sequential analysis, we pick five exemplary definitions of support and estimate the unexplained gender pay gap on support, $\delta_{S=1}$. Specifically, we restrict the sample to the employees on support and implement different estimators for the unexplained gender pay gap, which we explain in the next section. 

\subsection{Estimators}\label{est}
There exists a very large set of possible parametric and semi-parametric estimators, doubly-robust mixtures between the two types of estimators, and non-parametric estimators. We focus on common estimators that are easy to implement with the objective of showing the main trade-offs in estimator choice. We apply each estimator with five different support conditions where we restrict the sample to the observations satisfying the respective support definition.

\subsubsection{Dummy regression (LRM)} 
A simple estimation approach for the unexplained gender pay gap employs a linear regression model (LRM) with a dummy for women,
\begin{equation} \label{lm1}
Y_i = \alpha + G_i {\delta}_{LRM}  +  X_i{\beta} + {\varepsilon}_i,
\end{equation}
where $\alpha$ is a constant, the vector ${\beta}$ describes the association between the wage determinants $X_i$ and the wages, and ${\varepsilon}_i$ is an error term. The parameter ${\delta}_{LRM}$ can be interpreted as the unexplained gender pay gap.

The LRM imposes the restriction that the coefficients of the wage determinants as well as the unexplained gender pay gap are homogeneous across all individuals. This includes the assumption that the unexplained gender pay gap is equal for women and men. This modelling assumption is unnecessarily restrictive, because it can be relaxed by including interaction terms between gender and the wage determinants, which we discuss in the next section in the context of the BO decomposition.

\subsubsection{Blinder-Oaxaca (BO) decomposition}

The BO decomposition is a two-step estimator that allows for heterogeneous unexplained gender pay gaps \citep{BLI,OX}. In the first step, we estimate in the subsample of men the linear wage model
\begin{equation} \label{bo0}
Y_i = \alpha_0 + X_i{\beta}_0 + {u}_i,
\end{equation}
where $\alpha_0$ is an intercept and ${u}_i$ is an error term. The coefficients ${\beta}_0$ describe the association between the wage determinants $X_i$ and the wages of men. In the second step, we use the estimated coefficients from this regression to predict the counterfactual male wage $\hat{\mu}_0(X_i)\equiv \hat{\alpha}_0 + X_i\hat{\beta}_0$ for each woman in the sample and estimate the mean unexplained gender pay gap for women using
\begin{equation} \label{bo1}
\hat{\delta}_{BO} = \frac{1}{N_1} \sum_{i=1}^{N} G_i (Y_i - \hat{\mu}_0(X_i) )
\end{equation}
with $N_1 = \sum_{i=1}^{N} G_i$ (hats indicate estimated parameters). Note that this estimation procedure is numerically identical to using a linear prediction for female wages, $Y_i = \alpha_1 + X_i{\beta}_1 + {v}_i$, in the sample of women and calculating the unexplained part of the pay gap as 
\begin{equation*}
\hat{\delta}_{BO} =(\hat{\alpha}_1 - \hat{\alpha}_0) + \frac{1}{N_1} \sum_{i=1}^{N} G_i X_i(\hat{\beta}_1-\hat{\beta}_0),
\end{equation*}
since $\displaystyle \frac{1}{N_1} \sum_{i=1}^{N} G_i Y_i = \hat{\alpha}_1 + \frac{1}{N_1} \sum_{i=1}^{N} G_i X_i\hat{\beta}_1$.\bigskip

In contrast to the LRM, this version of the BO decomposition allows for gender differences in the impact of characteristics $X_i$ on wages as well as for heterogeneity in the unexplained gender pay gap that is driven by these differences. 

The BO model corresponds to the LRM augmented with interaction terms between gender and all observable wage determinants:
\begin{equation} \label{bo2}
Y_i = \alpha_0 + G_i  \underbrace{(\alpha_1-\alpha_0)}_{=\alpha_{BO}}  +  X_i{\beta_0} + G_i  X_i  \underbrace{(\beta_1-\beta_0)}_{= \beta_{BO}}  + {\epsilon}_i.
\end{equation}
Using this fully interacted LRM, we could estimate the BO unexplained wage gap by
\begin{equation*}
\hat{\delta}_{BO} =\hat{\alpha}_{BO} + \frac{1}{N_1} \sum_{i=1}^{N} G_i X_i\hat{\beta}_{BO}.
\end{equation*}
This estimation procedure is numerically identical to \eqref{bo1} when we control for the same characteristics $X_i$ as in \eqref{bo0}.

%Alternatively, \cite{bach18} and \cite{imai13} propose machine learning approaches which allow to obtain unbiased estimates of $\alpha_{BO}$ and $\beta_{BO}$. However, we focus on (\ref{bo1}) to implement the Blinder-Oaxaca decomposition, since we are mainly interested in $\delta_{BO}$ and not in the structural parameters $\alpha_{BO}$ and $\beta_{BO}$.

\subsubsection{Inverse probability weighting (IPW) estimators}

Inverse probability weighting (IPW) estimators are also two-step estimators \citep{hi03,ho52}. In the first step, we estimate the conditional probability of being a women, the so-called propensity score, based on the model
\begin{equation*}
{p}(X_i) = {Pr}(G_i=1|X_i) = F( X_i{\gamma} ),
\end{equation*}
where $F(\cdot)$ is a binary link function (e.g., the logistic distribution). In the second step, we predict $\hat{p}(X_i)$ for all observations and estimate the re-weighted sample average
\begin{equation*}
\hat{\delta}_{IPW} = \frac{1}{N_1} \sum_{i=1}^{N} G_i Y_i -  \sum_{i=1}^{N} \hat{W}_i^{0} Y_i,
\end{equation*}
with the IPW weights 
\begin{equation}\label{ipw}
\hat{W}_i^{0} = \displaystyle \frac{(1-G_i)\hat{p}(X_i)}{1-\hat{p}(X_i)}\Big/\displaystyle \sum_{i=1}^{N} \frac{(1-G_i)\hat{p}(X_i)}{1-\hat{p}(X_i)}
\end{equation}
The denominator in \eqref{ipw} guarantees that the IPW weights add up to one in finite samples \citep[see, e.g.,][]{bus14}. In contrast to LRM and BO, IPW does not impose any specific functional form on the relationship between $X_i$ and the wage, and it does not restrict heterogeneity in unexplained gender pay gaps. A potential disadvantage of IPW is its instability when the conditional probability of being a women is close to one \citep[e.g.,][]{ka10}, which may results in high variance of $\hat{\delta}_{IPW}$. We impose trimming rules to avoid propensity score values that are too extreme \citep[see the discussion in, e.g.,][]{str19}. In particular, we omit males with large weights $\hat{p}(X_i)/(1-\hat{p}(X_i))$ above the 99.5\% quantile.\footnote{We apply the normalization described in \eqref{ipw} after the trimming.} We document the number of trimmed observations in Table B.1 of Online Appendix B.

\cite{kline11} argues that BO estimators are equivalent to propensity score re-weighting estimators, but in contrast to IPW they model the propensity score linearly.\footnote{The usual practice of IPW estimators is to use a Logit or Probit estimator for the propensity score}. As a result, the implicit weights of BO can become negative, which is not possible for the IPW estimator. However, in the limit to a fully saturated BO model, even the linear specification of the propensity score would be well behaved and negative weights would be unlikely. 

Augmented inverse probability weighting (AIPW) dates back to \cite{rob94} and has received significant attention since \cite{Chernozhukov2017} proposed this estimation procedure in the context of machine learning. AIPW is a doubly-robust mixture between the BO and IPW approach,
\begin{equation} \label{ipw2}
\hat{\delta}_{AIPW} = \frac{1}{N_1} \sum_{i=1}^{N} G_i (Y_i - \hat{\mu}_0(X_i) ) -  \sum_{i=1}^{N} \hat{W}_i^{0} (Y_i - \hat{\mu}_0(X_i) ),
\end{equation}
with $\hat{\mu}_0(X_i)$ as in BO. The first right-hand term in \eqref{ipw2} is equivalent to the BO in \eqref{bo1}. The second right-hand term in \eqref{ipw2} has an expected value of zero, but makes a finite sample adjustment by re-weighting the observable bias of $\hat{\mu}_0(X_i)$ in the sample of men with the IPW weights. 

AIPW is more robust to misspecification than BO or IPW. In particular, $\hat{\delta}_{AIPW}$ is consistent even when either $\hat{\mu}_0(x)$ or $\hat{p}(x)$ is misspecified.
Moreover, the theoretical properties of AIPW are well established for generic estimators of the nuisance parameters $\hat{\mu}_0(x)$ and $\hat{p}(x)$.
%\footnote{In contrast, the theoretical properties of the PDS procedure of \cite{bell13} rely on the LASSO.} 
In particular, $\sqrt[4]{N}$-consistency of $\hat{\mu}_0(x)$ and $\hat{p}(x)$ is sufficient to achieve $\sqrt{N}$-consistency of $\hat{\delta}_{AIPW}$ (in combination with the cross-fitting procedure described below). This permits the application of flexible machine learning methods to estimate $\hat{\mu}_0(x)$ and $\hat{p}(x)$, which often have a convergence rate below $\sqrt{N}$. AIPW combined with machine learning is often called double-machine-learning \citep[see][for a review]{kna20b}. 

\subsubsection{Matching estimators}\label{match}

Exact matching (EXM) as used, for example by \citet{nopo08}, is a fully non-parametric estimation approach. We stratify the sample into $K \ll N$ mutually exclusive groups $W_i \in \{1, ..., K\}$, which are defined based on the characteristics $X_i$. The EXM estimator is
\begin{equation*}
\hat{\delta}_{EXM} = \frac{1}{N_1} \sum_{i=1}^{N} G_i \left(  Y_i - \sum_{j=1}^{N} \frac{\displaystyle  1\{W_i= W_j\} (1-G_j)Y_j}{\displaystyle \sum_{j=1}^{N} 1\{W_i= W_j\} (1- G_j)}  \right). 
\end{equation*}
EXM is more flexible than LRM, BO, IPW and AIPW. However, it suffers from the curse of dimensionality. When we consider many strata $K$, the risk of lacking support (i.e., finding no men matching specific strata of women) increases. Unlike the previous estimators, EXM cannot extrapolate into regions without support. Therefore, the estimation breaks down in the presence of support violations. Furthermore, even when there is support, there may be strata with very few men, which might lead to high variance of $\hat{\delta}_{EXM}$. Despite these potential disadvantages, our large data set with more than one million observations remains suitable for EXM. However, we have to define relatively coarse groups for the two continuous variables of age and tenure to avoid empty strata. For the same reason, we have to combine some industries and occupations with only few observations into more loosely similar groups. We implement EXM as in \citet{nopo08}; i.e., we exactly match the stratified variables that define common support. This implies that the versions with less restrictive support definitions also match on fewer variables, thus removing fewer differences in observed wage determinants. In the private [public] sector, we have 140 [142] strata for the least restrictive support definition and 844,100 [208,860] strata for the most restrictive one. 

We consider two additional semi-parametric matching estimators to account for potential concerns regarding the curse of dimensionality and unmatched wage determinants within strata. First, we use propensity score radius matching (PSM) \citep[see e.g.][]{f07,lw09,lmw11}. The propensity score $p(X_i)$ is the same as for IPW. PSM is a one-dimensional matching approach (as opposed to the multi-dimensional matching of EXM). To each woman, we match the men who have a propensity score value within a certain absolute difference (radius) of the woman's propensity score value. We define the radius as the 99\% quantile of the distribution of the closest absolute distances for all women, then omit women without a match within the radius. Table B.2 in Online Appendix B documents the number of females we use for the different matching estimators. We lose only a relatively few women who lack matching men within the defined radius.  

Second, we consider mixtures between exact and propensity score radius matching (EXPSM). Here, we exactly match on the wage determinants that define support, as in EXM, and apply propensity score radius matching within each stratum using all wage determinants, as in PSM. In contrast to EXM, the propensity score corrects semi-parametrically for remaining within-strata observable differences between women and men. In contrast to PSM, we enforce exact comparability of women and men with respect to some wage determinants.\footnote{It should be noted that we implement PSM after matching exactly on the three variables that define the least restrictive Support 1, because this improves matching without lack of support. As a result, PSM and EXPSM are identical for Support 1 but not for the other support definitions.}

\subsection{Flexibility of inclusion of wage determinants \label{nui}}

To estimate the so-called nuisance parameters, i.e., the wage equation $\hat{\mu}_0(x)$ and the propensity score $\hat{p}(x)$, we consider three different models. We call them the baseline, full, and machine learning (ML) models. The baseline and full models include the same wage determinants, but we vary the flexibility in terms of non-linear (e.g., polynomials, categorical dummies) and interaction terms of the wage determinants. The ML model can select the relevant wage determinants and the model flexibility in a data-driven way, using the full model specification as input. In practice, it is often unclear which non-linear and interaction terms are relevant. Using a too parsimonious model could bias ${\delta}$, due to model misspecification. Allowing for too much flexibility could lead to imprecise estimates of ${\delta}$ with high variance, especially in smaller samples. Accordingly, we face a bias-variance trade-off.

\subsubsection{Manual model specification}

In the baseline model, we control for all observed variables in a standard way. Specifically, we include age (linear and squared), tenure (linear and squared), vocational education (9 categories), citizen status (6 categories), marital status (3 categories), occupation (39 [20] categories in the private [public] sector), industry sector (36 [12] categories in the private [public] sector), management position (5 categories [and a missing dummy for public sector]), region (7 categories), establishment size (5 categories), and employment share as a fraction of full time (4 categories). Furthermore, we control for the dummy variables of temporary employment, employment contract with hourly wages, collective wage agreement, overtime hours payment, bonus payments (e.g., from profit sharing), supplementary wages (e.g., for shift work), and extra salary (e.g. Christmas and holiday salaries). Overall, the baseline model includes 117 control variables in the specification for the private sector and 74 control variables in the specification for the public sector.

In the full model, we add to the baseline model all non-linear and interaction terms that we think could potentially be relevant. The non-linear terms are the polynomials of age and tenure up to order seven, as well as four age and five tenure categories. Furthermore, we include interaction terms between control variables to allow for heterogeneous returns to important wage determinants.\footnote{We partially coarsen the categorical variables to improve support of the interaction terms.} Specifically, we interact occupation and industry groups with the categorical variables for age, tenure, education, foreigner status, management position, and establishment size. We also include interaction terms between establishment size categories and the categorical variables for age, tenure, vocational education, migration, and management position. Finally, we interact vocational education and management position categories with age and tenure categories. Overall, the full model includes 615 control variables in the specification for the private sector and 396 control variables in the specification for the public sector.

\subsubsection{Data-driven model specification}

In contrast to manual model specification, machine learning estimators have the potential to balance the bias-variance trade-off in a data-driven way. We use the LASSO to specify our ML models. The LASSO is a penalized regression method \citep[see][for a detailed description of the LASSO]{hast09}. It can balance the bias-variance trade-off by shrinking some coefficients towards zero. Eventually, some covariates are excluded from the model when the coefficients are exactly zero. Accordingly, the LASSO is a model selection device. To obtain the predicted wages of men $\hat{\mu}_0(x)$ in the baseline and full models for BO, we estimate in the sample of men the OLS model
\begin{equation*} \label{eq3}
(\hat{\alpha}_0,\hat{\beta}_0)=\arg\min_{a,b} \frac{1}{N} \sum_{i=1}^{N}( Y_i - a- X_i b)^2,
\end{equation*} 
and predict the conditional male wages by $\hat{\mu}_0(x)= \hat{\alpha}_0 + x \hat{\beta}_0$.
%\footnote{For the PDS procedure, we estimate the wage equation using the entire sample and not only the sample of males \citep[see][for details]{bell13}.}
In contrast, the LASSO objective function is
\begin{equation} \label{eq3x}
(\hat{\alpha}_0,\hat{\beta}_0)=\arg\min_{a,b}\frac{1}{N} \sum_{i=1}^{N} (Y_i - a - X_i b )^2 + \lambda \|b \|_1,
\end{equation}
with the penalty term $\lambda \|b \|_1 = \lambda \sum_{p=1}^{P} |b|$. The tuning parameter $\lambda\geq 0$ specifies the degree of penalization of the absolute coefficient size of $\hat{\beta}_0$. If $\lambda=0$, the ML model is equivalent to the full model.
If $\lambda>0$, some coefficients $\hat{\beta}_0$ shrink towards zero and eventually approach zero exactly. Shrinking coefficients to zero is equivalent to excluding the corresponding control variable in $X_i$ from the wage equation. In the extreme case, when $\lambda \rightarrow \infty$, all coefficients in $\hat{\beta}_0$ are exactly zero and only the intercept $\hat{\alpha}_0$ is non-zero, because it is not penalized. We determine the tuning parameter $\lambda$ with a 5-fold cross-validation procedure \citep{chet17}.\footnote{Alternatively, the choice of the tuning parameter $\lambda$ can be based on information criteria \citep{zou07} or data-driven procedures \citep{bell12}.} The cross-validation procedure selects the $\lambda$ that minimises the mean-squared-error (MSE). To smooth the ML model, we apply the one-standard-error rule to the minimum $\lambda$ \citep[see, e.g.,][]{hast09}. 

We use similar models to estimate the propensity score $\hat{p}(x)$. However, instead of using OLS, we use the Logit estimator for the baseline and full model. Likewise, we use Logit-LASSO instead of the standard LASSO approach for the ML model \citep[see][for an introduction to Logit-LASSO]{hast16}.\footnote{The only exception is the post-double-selection procedure, for which we use a linear model to estimate the propensity score, as proposed in \cite{bell13}.} In Figures C.1-C.8 in Online Appendix C, we report the density of the estimated propensity scores by gender. % Typically, the LASSO selects less control variables for the propensity score than for the wage equation (see Table \ref{cov} in Online Appendix C). 

\textbf{Post-double-selection (PDS) procedure.} The LASSO version of the LRM corresponds to the PDS procedure of \cite{bell13,bell14}. The conventional LASSO has the purpose of predicting $Y_i$, but it is not designed to obtain an unbiased estimate of the structural parameter ${\delta}_{LRM}$. %Naively applying LASSO (\ref{lm1}) can lead to biased and inconsistent estimates of ${\delta}_{LRM}$.
To illustrate this, the application of LASSO to (\ref{lm1}) without penalizing the gender dummy would most likely lead to the omission of those wage determinants that are highly correlated with gender, such that an omitted variable bias is inevitable. The PDS procedure enables LASSO to select the relevant wage determinants in a data-driven way, without the threat of omitting important variables. The PDS procedure has three steps. First, the LASSO selects the relevant control variables in the linear wage equation $Y_i = \alpha_Y +  X_i{\beta_Y} + {\nu}_i$, which does not contain the the gender dummy $G_i$.\footnote{In contrast to \eqref{eq3x}, we use the sample of men and women for the wage equation of the PDS procedure.} In practice, the LASSO can set some coefficients in the vector ${\beta_Y}$ to zero, which is equivalent to excluding the corresponding control variables in $X_i$ from the wage equation. Second, the LASSO selects the relevant dimensions of $X_i$ in the linear gender model $G_i = \alpha_G +  X_i{\beta_G} + {\eta}_i$. As before, the LASSO estimator can set some coefficients in the vector $\beta_G$ to zero, such that only those control variables with a strongly gender correlation remain in the model. We denote the union of all wage determinants with non-zero coefficients in either $\beta_Y$ or $\beta_G$ by $\tilde{X}_i$. Finally, we estimate the linear OLS model (called 'post-LASSO')
\begin{equation*} 
Y_i = \alpha_{PDS} + G_i {\delta}_{PDS}  +  \tilde{X}_i{\beta}_{PDS} + {\tilde{\varepsilon}}_i.
\end{equation*}
\cite{bell13} show that the estimator of ${\delta}_{PDS}$ is consistent and asymptotically normal. The PDS procedure allows for the inclusion of high-dimensional characteristics $X_i$ (i.e., more characteristics than observations), but requires $\tilde{X}_i$ to be sparse. Some approximation error in the selection of the $\tilde{X}_i$ variables is permitted. The PDS procedure is doubly robust to misspecification of either the earnings equation or the gender model.

For the same reason as for the LRM, applying LASSO in a naive way to the interacted BO in (\ref{bo2}) could lead to biased estimates of $\alpha_{BO}$ and $\beta_{BO}$, which could also bias $\hat{\delta}_{BO}$. However, we can use the LASSO to estimate $\hat{\mu}_0(x)$, since $\hat{\mu}_0(x)$ is a pure prediction model for $Y_i$, and then use $\hat{\mu}_0(x)$ as a plug-in estimator in \eqref{bo1}. Accordingly, \eqref{bo1} provides a way to combine the Blinder-Oaxaca decomposition with standard machine learning methods. This procedure is called T-learner in the denomination of \cite{kuenz19}. A possible disadvantage of this approach is that we do not obtain estimates of the parameters $\alpha_{BO}$ and $\beta_{BO}$, since they are not required to estimate $\delta_{BO}$. Alternatively, \cite{bach18} propose a procedure that is based on post-double-selection, but is more flexible and allows us to obtain estimates of selected parameters in $\alpha_{BO}$ and $\beta_{BO}$.

\textbf{Cross-fitting.} To avoid over-fitting and obtain $\sqrt{N}$ convergence for AIPW, it is necessary to estimate the ML models of $\hat{\mu}_0(x)$ and $\hat{p}(x)$ in a different sample than $\hat{\delta}_{AIPW}$ \citep[see][]{che17}. We achieve this by using cross-fitting. We partition the sample into two parts of equal size. We estimate $\hat{\mu}_0(x)$ and $\hat{p}(x)$ with the first partition, extrapolate their fitted values to the second partition and use these values to estimate $\hat{\delta}_{AIPW}$. Thereafter, we switch the first and second partition and repeat the procedure, such that the entire data set is used efficiently. We report the average $\hat{\delta}_{AIPW}$ across both partitions. 

%Selecting the model for the nuisance parameters and estimating $\hat{\delta}$ in the same sample could lead to over-fitting \citep[see formal arguments in][]{che17}. To overcome this potential disadvantage, we additionally use a cross-fitting procedure. This implies we partition the data in two samples. We use the first sample to select and estimate the nuisance parameters $\hat{\mu}_0(x)$ and $\hat{p}(x)$. We extrapolate the fitted values to the second sample and estimate $\hat{\delta}$ in the second sample. To use the data efficiently, we switch the first and second sample and repeat the cross-fitting procedure. Finally, we report the average of both $\hat{\delta}$ estimates.

\textbf{Selected variables.} We allow the LASSO to select among all control variables of the full model. Table C.1 in Online Appendix C documents the number of selected control variables for all ML models we consider. The final specifications vary by support and  estimation procedure. For the wage equation, the ML model considers between 371 and 513 control variables in the different specifications for the private sector and between 215 and 306 control variables for the public sector. For the propensity score model, the ML model considers between 141 and 428 control variables for the private sector and between 126 and 230 control variables for the public sector. 

\textbf{Performance.} In Tables C.2 and C.3 in Online Appendix C, we report the out-of-sample prediction power of the different nuisance parameter models. To obtain the out-of-sample prediction power, we use a cross-fitting procedure.\footnote{We partition the data in two equally sized samples, using one partition as a training sample and the other as a test sample. Thereafter, we switch the two partitions and report the average prediction power across the two samples.} The prediction power of the baseline, full, and ML models does not differ strongly. Accordingly, the baseline model already explains a significant amount of the variation in the data. The prediction power of the full model is systematically better than the prediction power of the baseline model, but the additional gain is moderate. The ML model cannot systematically outperform the full model. The prediction power of the full and ML models is fairly similar, even though the ML model controls for many fewer variables than the full model. This suggests that the full model is too flexible. However, because of our large data set, the prediction power of the full model does not deteriorate compared to the ML model. The degrees of freedom loss of the full compared to the ML model appear to be of minor importance. However, we expect that the ML model would outperform the full model in smaller samples. 

\section{Results}

\subsection{Analysis of common support}\label{sec:sup}
Our first step is to analyze common support using the approach of \citet{nopo08}. We sequentially increase the number of wage determinants for which we impose common support. To determine the order in which we add variables, we run a simple wage regression in the sample of men, i.e. we estimate $\mu_0(x)$ in order to determine the importance of the different variables for explaining men's wages. We measure importance as the change in adjusted $R^2$ when we omit one block of variables, e.g.\ all occupation dummies, but keep all other wage determinants. We do this separately for the private and public sectors and sort all variable blocks in decreasing order according to the average $R^2$ changes across both sectors. Thus, we start with the most important variable block for explaining male wages and finish with the least important. We report the resulting order of variables and the sector-specific changes in the adjusted $R^2$ in Tables B.3 and B.4 of Online Appendix B for the private and public sector, respectively.
 
Our choice of ordering based on importance for explaining male wages is motivated by the curse of dimensionality. The larger the number of variables with respect to which we enforce common support, the more likely it is that there will be empty cells. As a result, we expect the share of the original sample for which there is support to decrease as we add more and more variables. This implies that there is some difficulty in achieving a reasonable balance of comparability, sample size and representativeness of the remaining sample. Therefore, we want to impose support with respect to variables that significantly influence the counterfactual male wage, but not necessarily with respect to variables of minor importance. 

\begin{figure}[htbp]
\caption{Analysis of common support by sector} \label{cs_graph}
\begin{center}
\begin{subfigure}{0.8\textwidth}
\includegraphics[width=\textwidth]{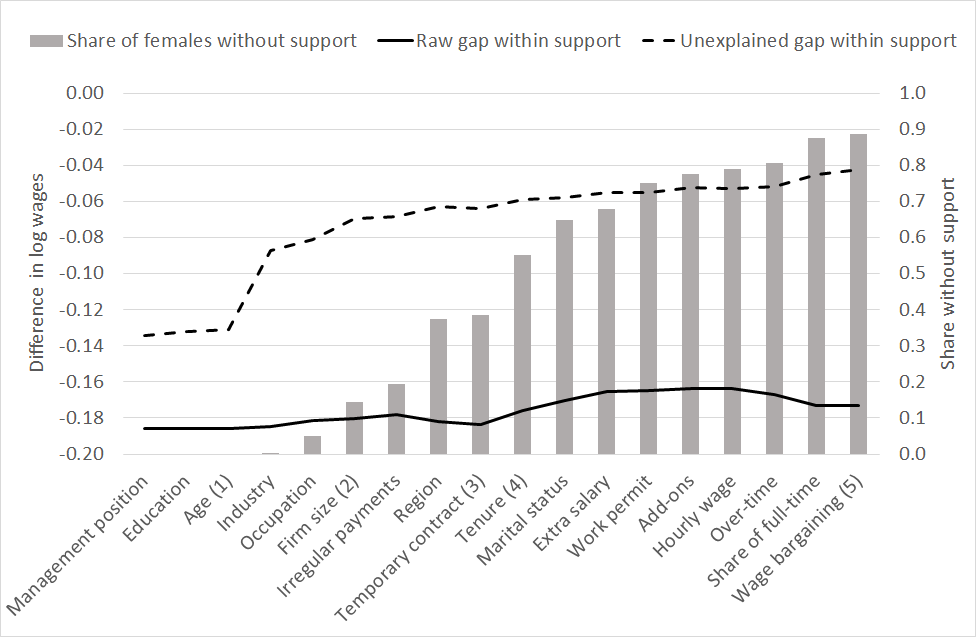}\caption{Private sector}
\end{subfigure}
\begin{subfigure}{0.8\textwidth}
\includegraphics[width=\textwidth]{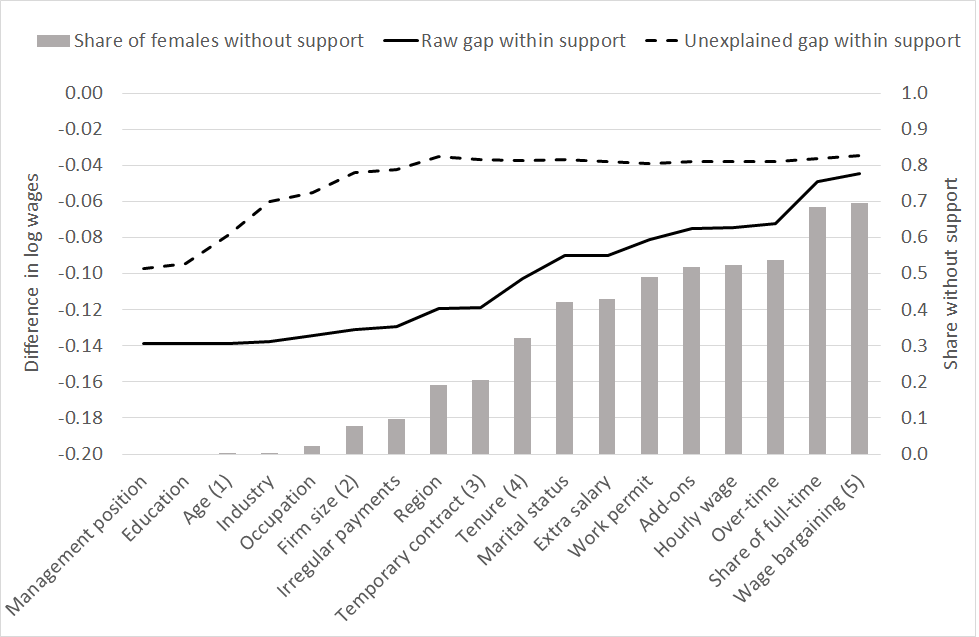}\caption{Public sector}
\end{subfigure}
\end{center}
\par
{\footnotesize Notes: The indicated wage determinants are added sequentially from left to right. The raw gap is the difference between the average log wages of women and men in the sample with full support with respect to the indicated variables. The unexplained gap is based on the counterfactual male wage obtained from exact matching on all wage determinants from the left to the respective determinant within the sample that has full support with respect to these variables. The numbers in parentheses indicate the five support definitions we use for the estimations.}
\end{figure}

In Figure \ref{cs_graph}, we report how the share of women with support falls as we increase the number of wage determinants with respect to which we enforce support. There is full support with respect to the three most important determinants -- management position, education and age -- in the private sector, and only 2 women are off support in the public sector. The loss of observations due to lack of support is relatively small when we add industry and occupation, which are both highly relevant for predicting wages. It becomes larger but remains below 20\% if we add establishment size and a dummy for irregular payments. Thereafter, however, support deteriorates quickly. Once we enforce support with respect to the 10 variables that are most important for explaining male wages, 55\% of women in the private sector and 32\% of women in the public sector have no comparable men with respect to these variables. The loss of women is stronger in the private sector than in the public sector because the latter is more homogeneous in terms of occupations and industries. Subsequently adding the less important wage determinants further reduces support, resulting in lack of support for 89\% of women in the private sector and 70\% of women in the public sector when we enforce full support with respect to all observed wage determinants.

Of course, different orders of variables lead to different evolutions of support. Table B.5 in Online Appendix B shows how support changes when we add these variables in three alternative orders. The first one uses \textit{sector-specific} $R^2$ changes rather than the average over both sectors to order variables. For the main analysis, we want to keep the results comparable across sectors, which is why we want the same order for both sectors. The order resulting from \textit{sector-specific} $R^2$ changes differs across sectors for some variables, but the overall differences are moderate. Accordingly, the evolution of support when adding variables is also similar. The second alternative is a random order. Here support remains higher for a larger number of added variables, mainly because critical variables such as industry and tenure are, by chance, added relatively late. This would change with a different random order in which the are added earlier. The last alternative uses an \textit{increasing} order according to the sector-average $R^2$ changes as the other extreme to a decreasing order. Here, support breaks down rather late. The reasons for this are twofold. First, the ordering prioritizes wage determinants in which women and men do not greatly differ. Second, many of these wage determinants are dummies while more important variables (such as education, occupation and industry) have many different categories. This mechanically splits the sample in more cells than a dummy variable such that cell size is reduced.

Besides studying the evolution of support, we also estimate the elements of (\ref{cs2}). The raw gender pay gaps on support, $\Delta_{S=1}$ , can be obtained from the average gender difference in wages in the sample with support. To estimate the unexplained gender pay gap for the women with support, $\delta_{S=1}$, we follow \citet{nopo08} and calculate the pay gap that remains after matching exactly on all variables for which we enforce support. Thus, when adding support-relevant wage determinants, we also increase the number of variables on which we exactly match. The explained gender pay gap on support, $\eta_{S=1}$, can be calculated as the difference between the raw and unexplained gender pay gap within support. We report the main insights from this analysis in Figure \ref{cs_graph} and we document the full set of results in Tables B.3 and B.4 in Online Appendix B.

The raw gender pay gap is relatively insensitive to the support definition in the private sector. It decreases only slightly, from 18.6\% under the weakest support definition to 17.3\% under the strongest one. In contrast, the EXM estimates of the unexplained and explained gender pay gaps are very sensitive to the support definition. The unexplained gender pay gap on support falls from 13.4\% under the weakest support definition to 4.2\% under the strongest one. Correspondingly, the explained gender pay gap increases from 5.1\% to 13.1\% as we exactly match on more wage determinants. Interestingly, the results differ substantially for the public sector. Here, a substantial part of the raw gender pay gap is due to women without support. The raw gender pay gap on support decreases from 13.9\% under the weakest support definition to only 4.5\% under the strongest one. The unexplained gender pay gap on support decreases quickly from 9.7\% to to 3.5\% when we impose support with respect to, and exactly match on, the eight most important wage determinants. Thereafter, imposing stronger support and matching on more variables leaves the unexplained gender pay gap almost unchanged although the raw gender pay gap decreases. This finding is in line with strongly regulated wage setting of the Swiss public sector, which leaves little room for gender pay differences. 

For the analysis of the unexplained gender pay gap with different estimation methods and model flexibility, we pick five definitions of support that illustrate the trade-offs between comparability, sample size and representativeness. We indicate the five versions with the respective number in parentheses in Figure \ref{cs_graph}. Here it is important to note that for a given choice of variables it does not matter in which order support is enforced. All possible orders will lead to exactly the same sample. Support 1 does not change the sample except for excluding two women in the public sector. Thus, it effectively resembles the case without support enforcement. Support 2 enforces support with respect to classical and very important wage determinants. Support 3 covers all quantitatively important wage determinants. Support 4 adds tenure as the only proxy for actual experience, but it does not significantly explain wages. Support 5 is the most extreme case, which enforces full support with respect to all variables. Table B.6 in Online Appendix B shows how imposing stricter support changes the average characteristics of the remaining women with support. The changes are largest with respect to occupation, management position, establishment size, and part-time employment status. This is important for the interpretation of our results. In the presence of heterogeneous pay gaps, we cannot expect estimates to be the same when the sample changes due to stricter enforcement of support.

\subsection{Analysis of the unexplained gender pay gap}\label{sec:res_main}

\begin{figure}[]
\caption{Mean unexplained gender pay gap in the private sector} \label{results_graph_priv} 
\begin{center}
\begin{subfigure}{0.45\textwidth}
\includegraphics[width=\textwidth]{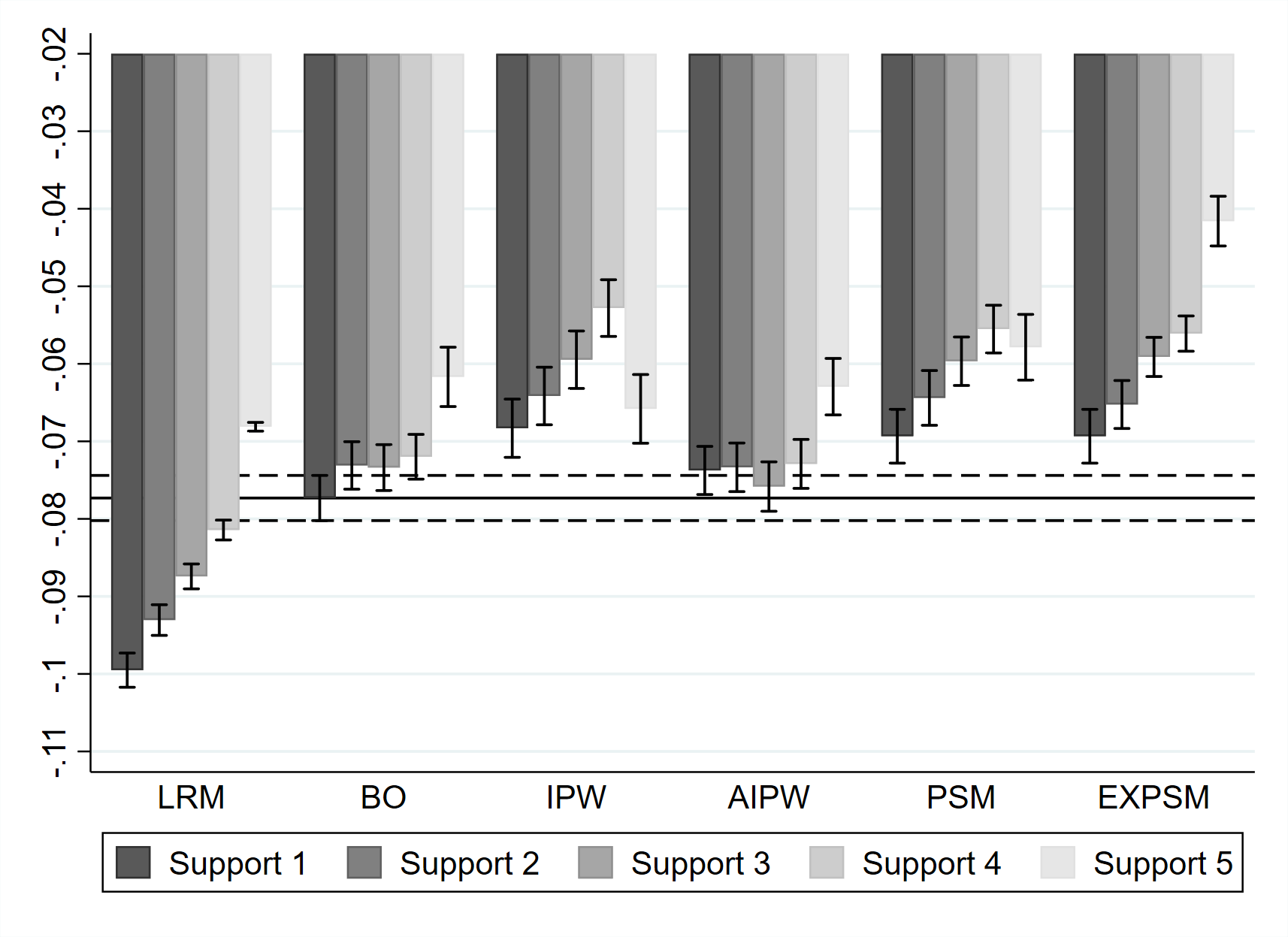}
\caption{Baseline model: estimates}
\end{subfigure}
\begin{subfigure}{0.45\textwidth}
\includegraphics[width=\textwidth]{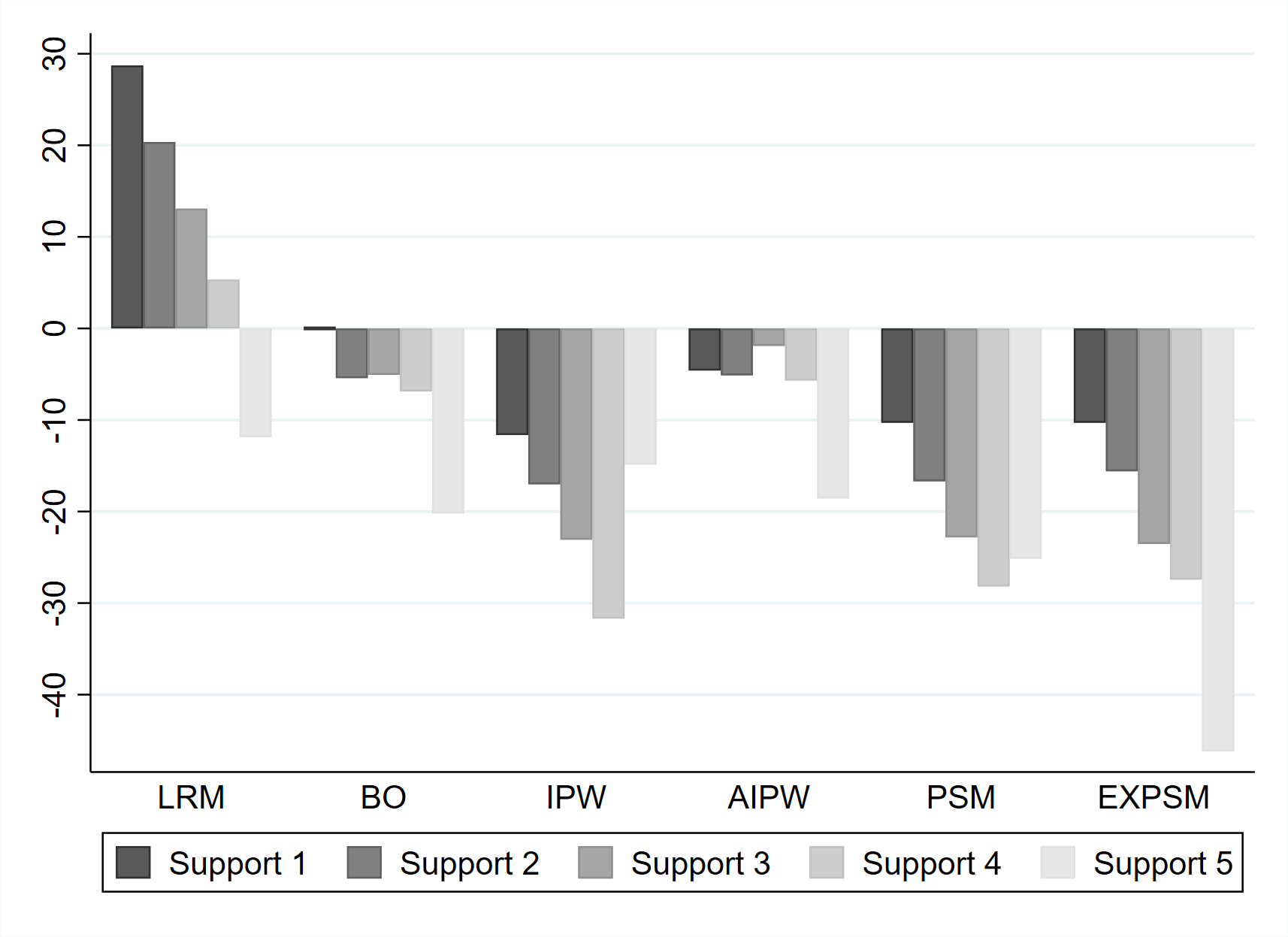}
\caption{Baseline model: difference to benchmark}
\end{subfigure}
\begin{subfigure}{0.45\textwidth}
\includegraphics[width=\textwidth]{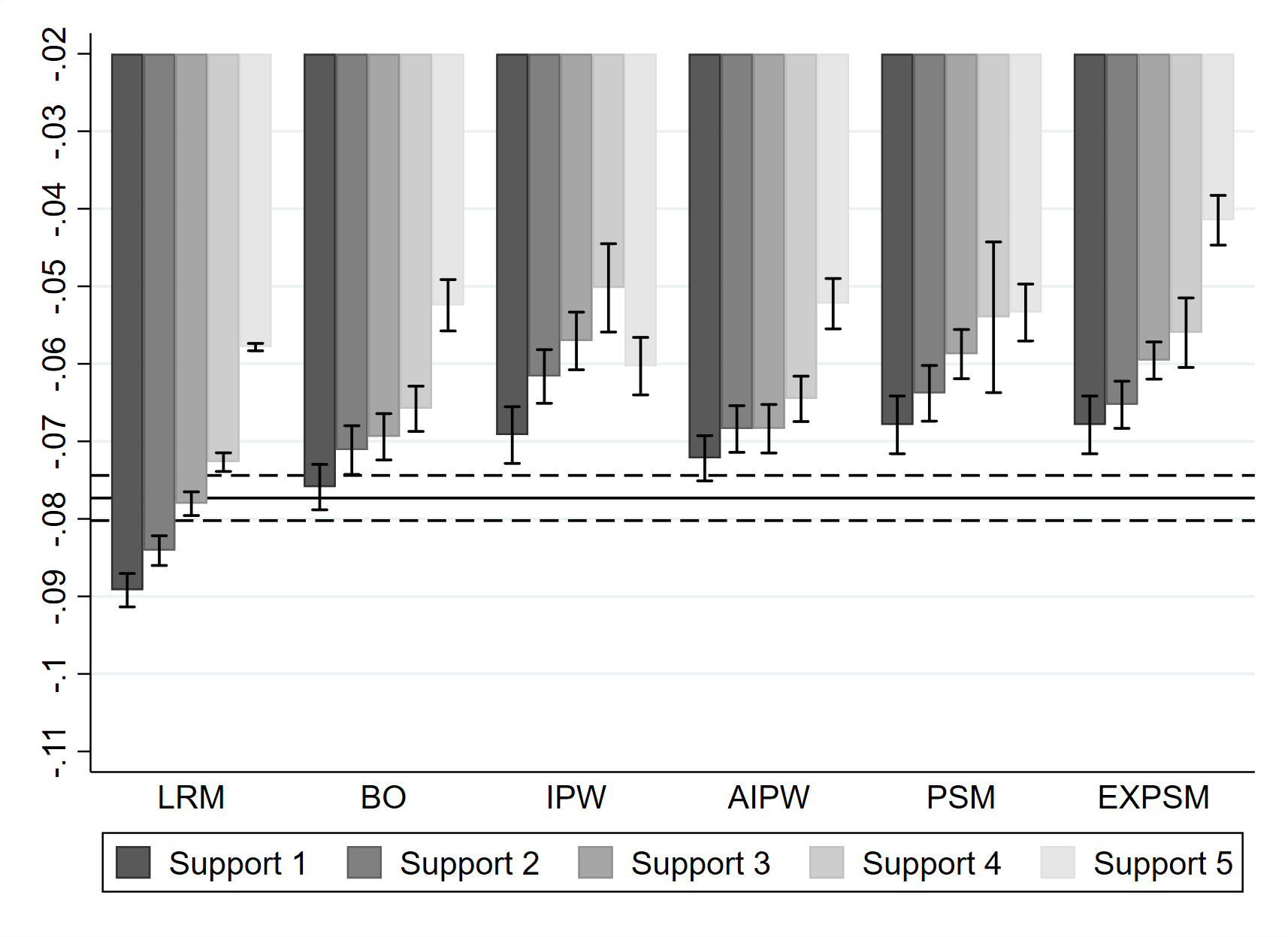}
\caption{Full model: estimates}
\end{subfigure}
\begin{subfigure}{0.45\textwidth}
\includegraphics[width=\textwidth]{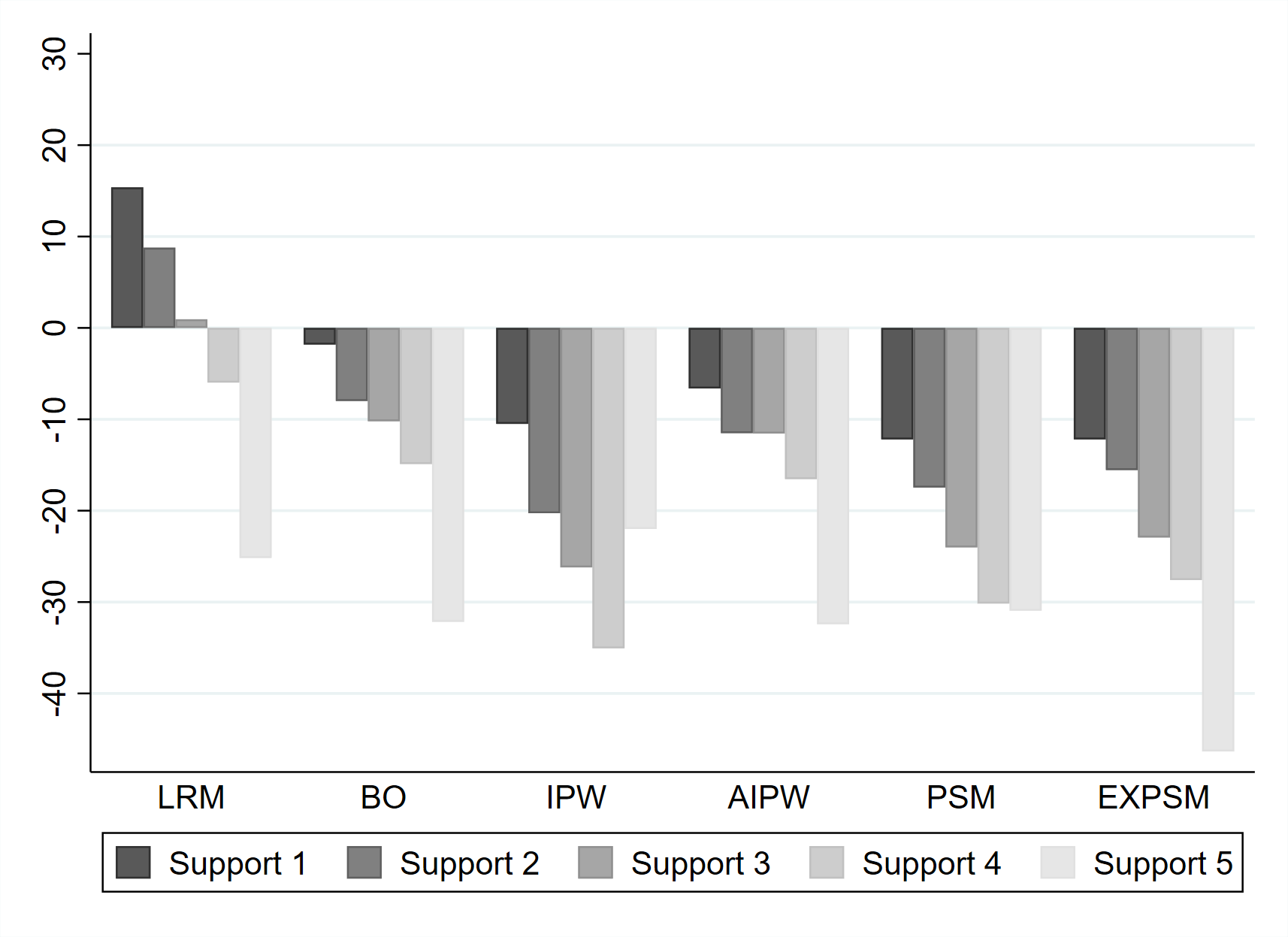}
\caption{Full model: difference to benchmark}
\end{subfigure}
\begin{subfigure}{0.45\textwidth}
\includegraphics[width=\textwidth]{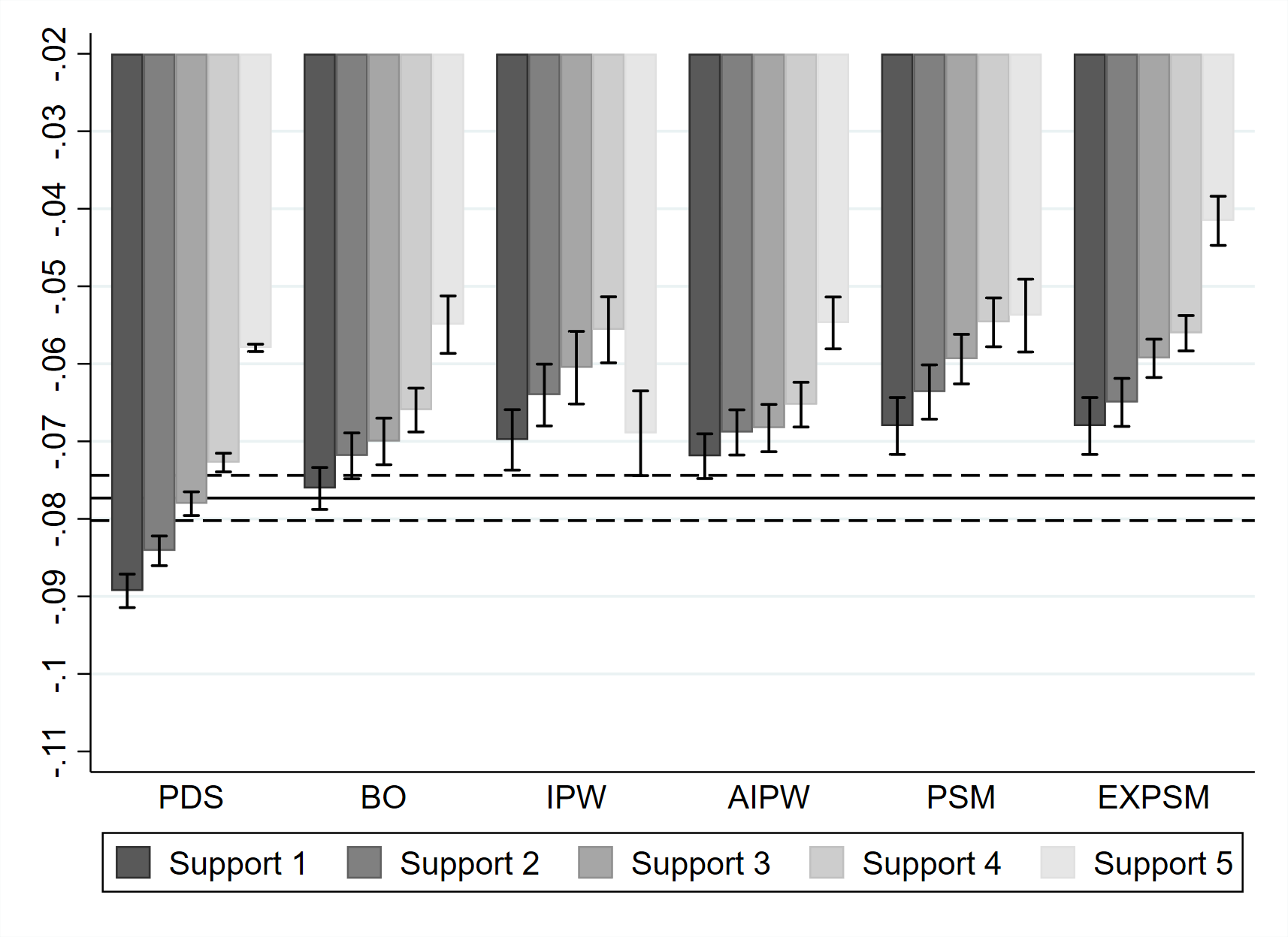}
\caption{ML model: estimates}
\end{subfigure}
\begin{subfigure}{0.45\textwidth}
\includegraphics[width=\textwidth]{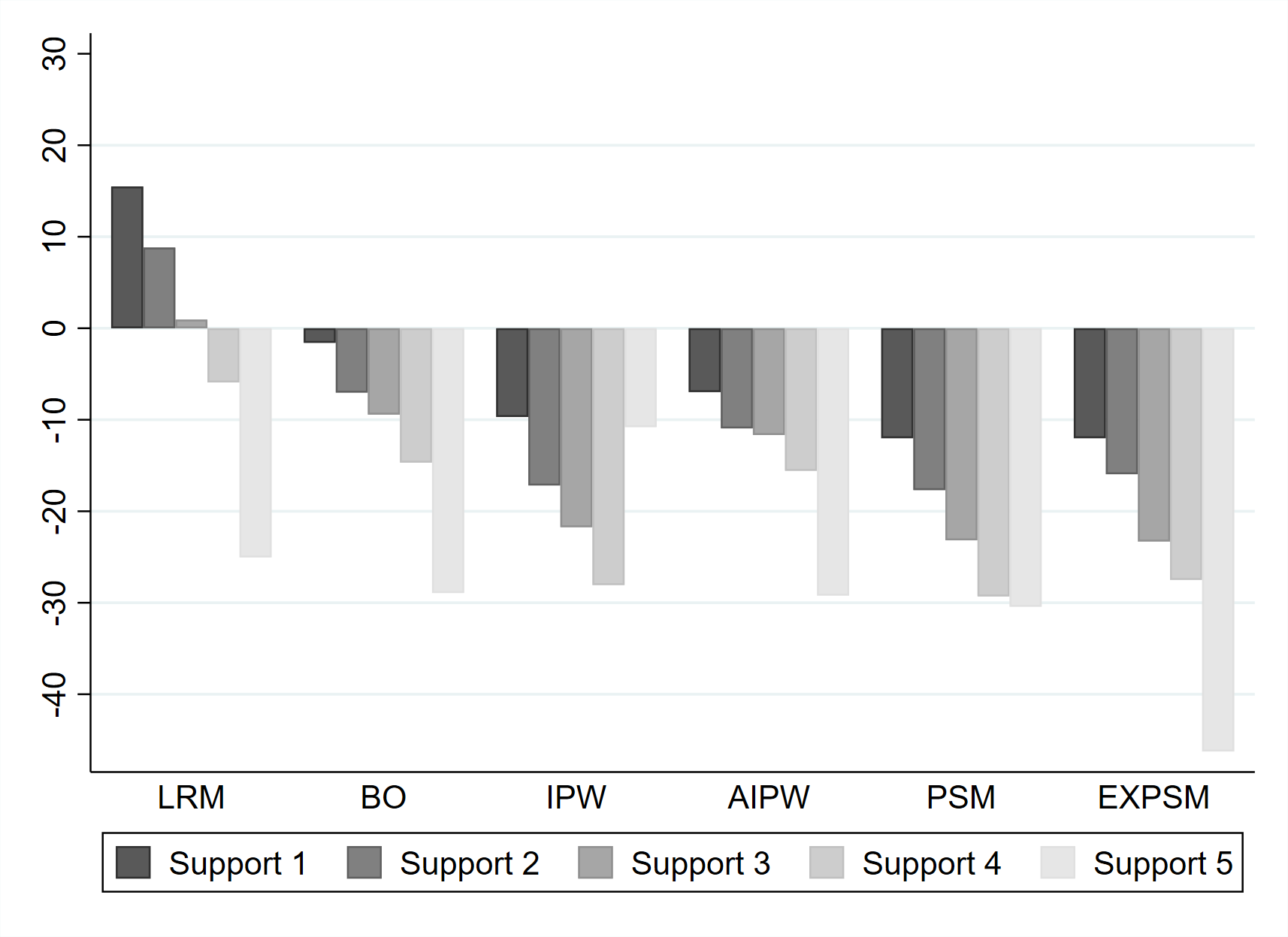}
\caption{ML model: difference to benchmark}
\end{subfigure}
\end{center}
{\footnotesize Notes: We report all estimates and their standard errors in Table D.1 in Online Appendix D. Capped lines indicate the 95\% confidence intervals of the estimated unexplained gender pay gap. The black horizontal line in the left panels shows the BO estimate of Support 1 as a benchmark with the baseline specification and its 95\% confidence band in black dashed lines. The right panels show the difference from this benchmark in percent, i.e., the difference from the BO estimate using Support 1 with baseline model specification. PDS is the abbreviation for post-double-selection procedure.}
\end{figure}

\begin{figure}[]
\caption{Mean unexplained gender pay gap in the public sector} \label{results_graph_pub} 
\begin{center}
\begin{subfigure}{0.45\textwidth}
\includegraphics[width=\textwidth]{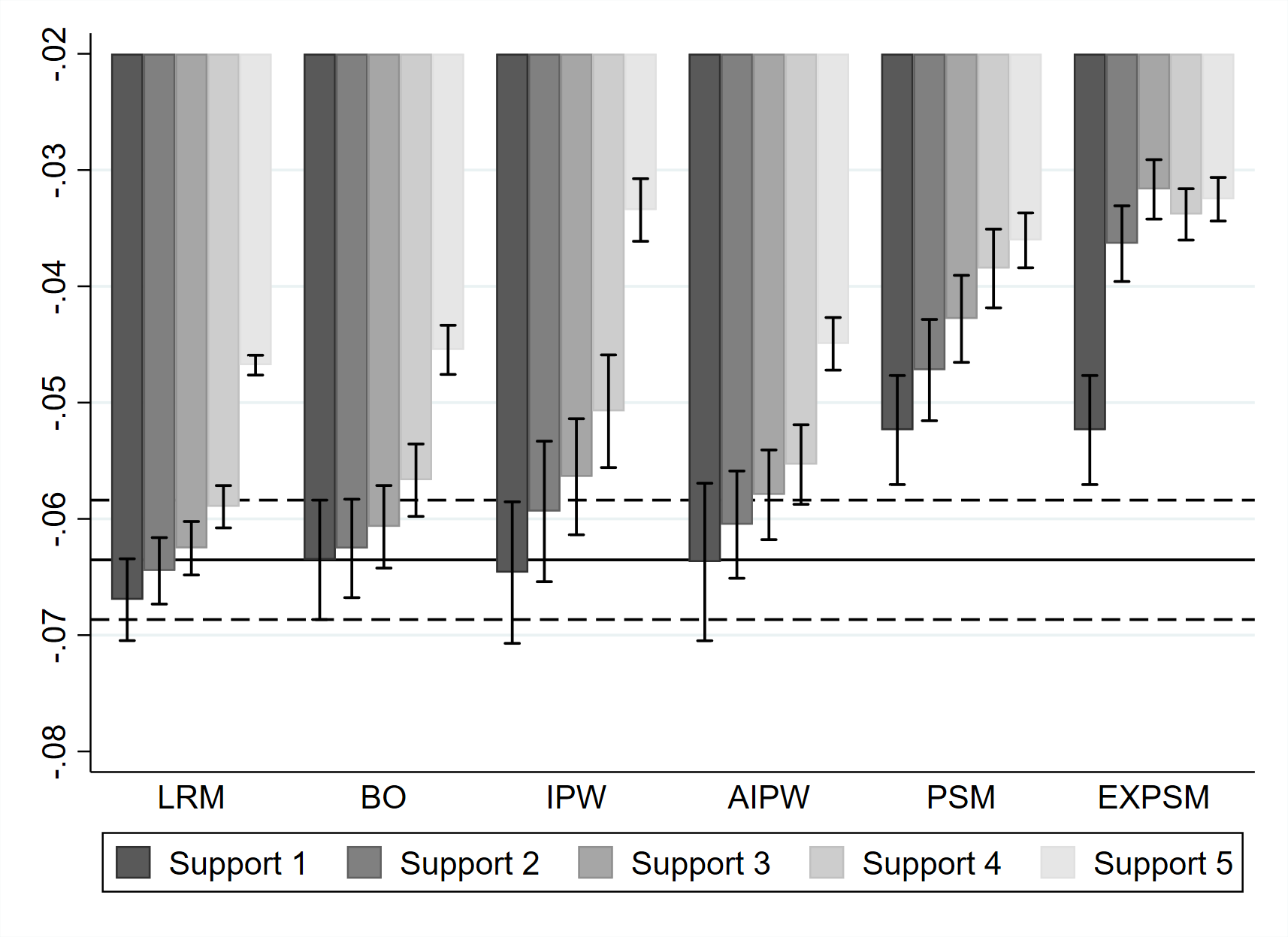}
\caption{Baseline model: estimates}
\end{subfigure}
\begin{subfigure}{0.45\textwidth}
\includegraphics[width=\textwidth]{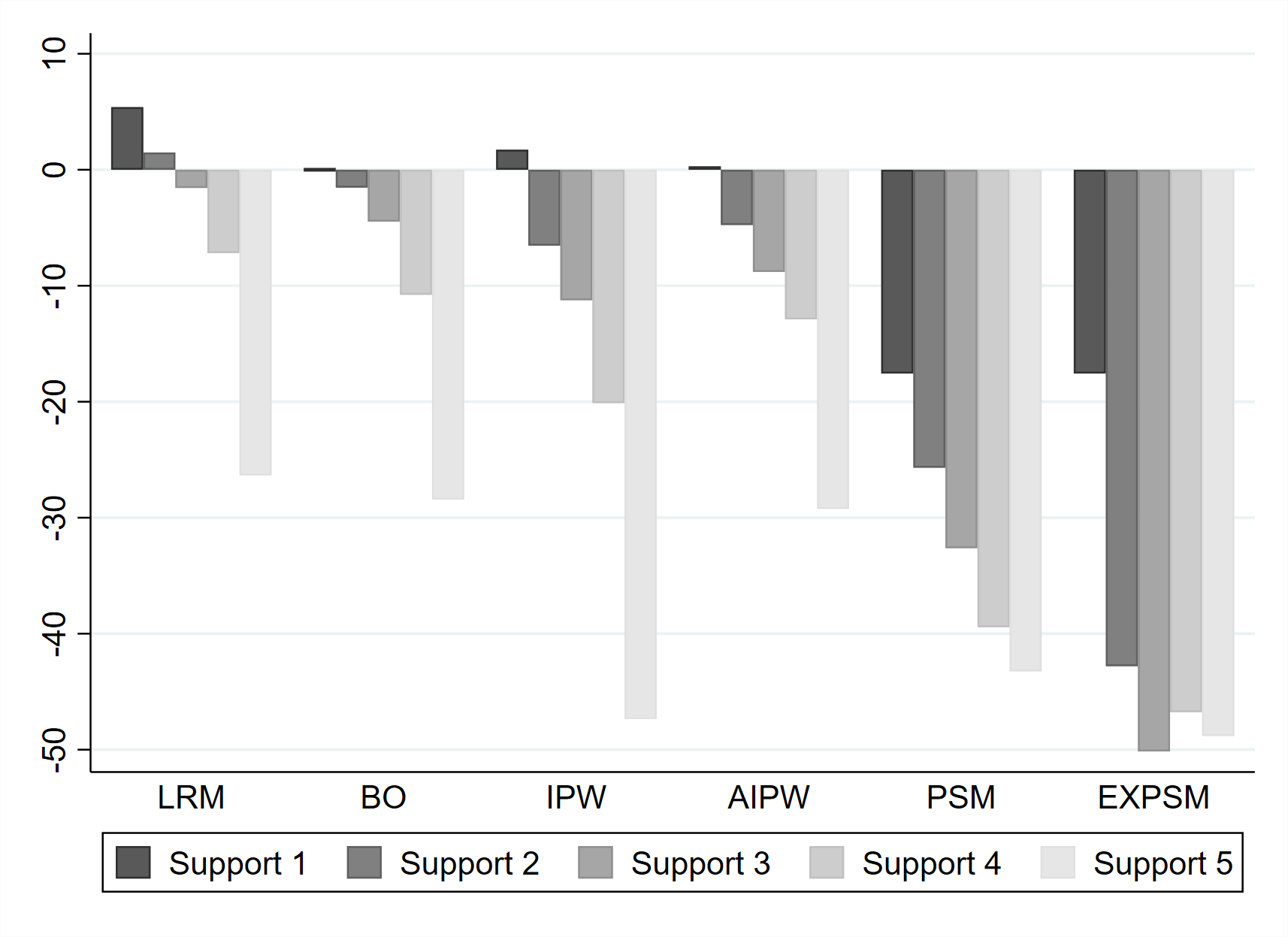}
\caption{Baseline model: difference to benchmark}
\end{subfigure}
\begin{subfigure}{0.45\textwidth}
\includegraphics[width=\textwidth]{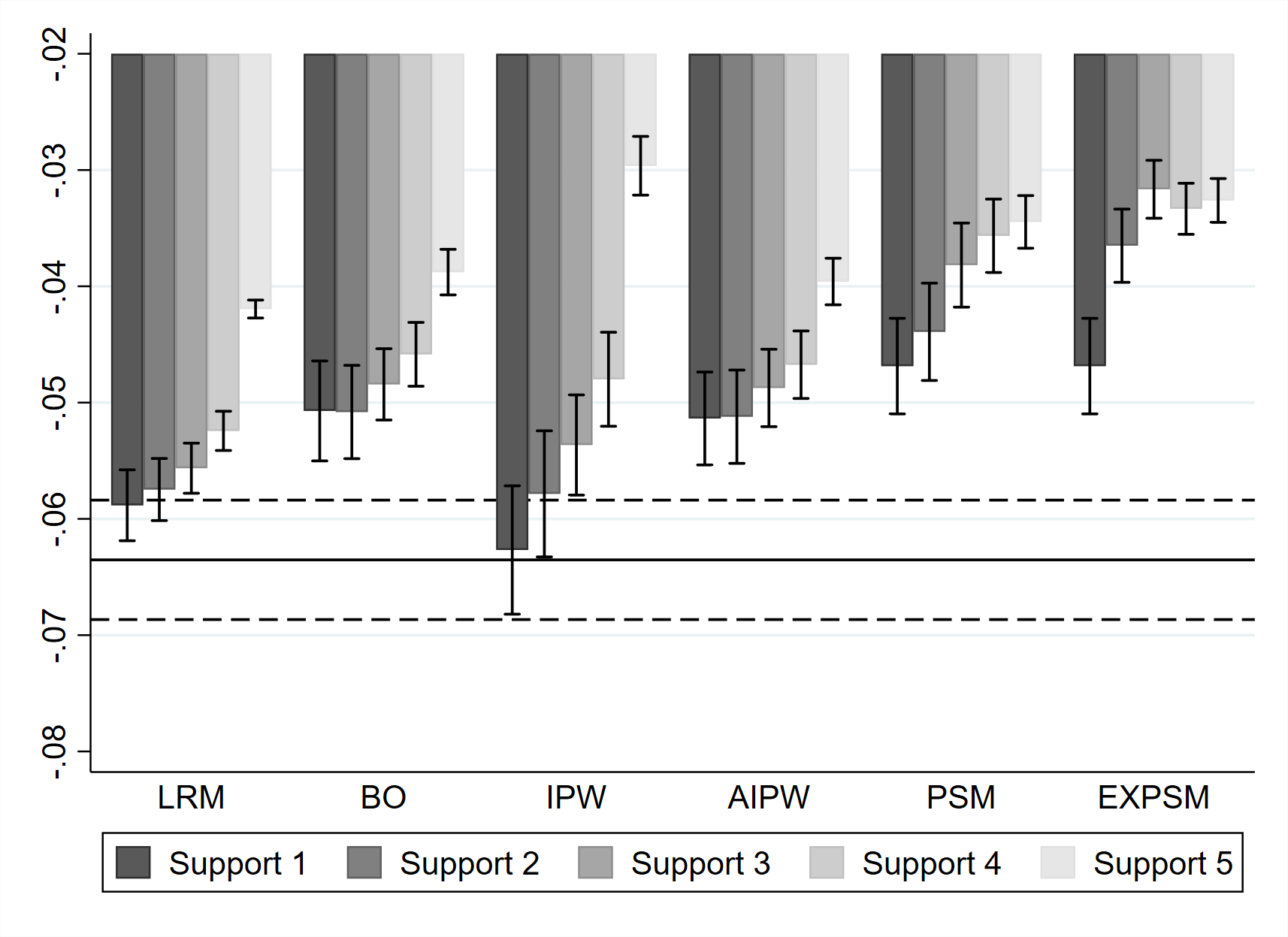}
\caption{Full model: estimates}
\end{subfigure}
\begin{subfigure}{0.45\textwidth}
\includegraphics[width=\textwidth]{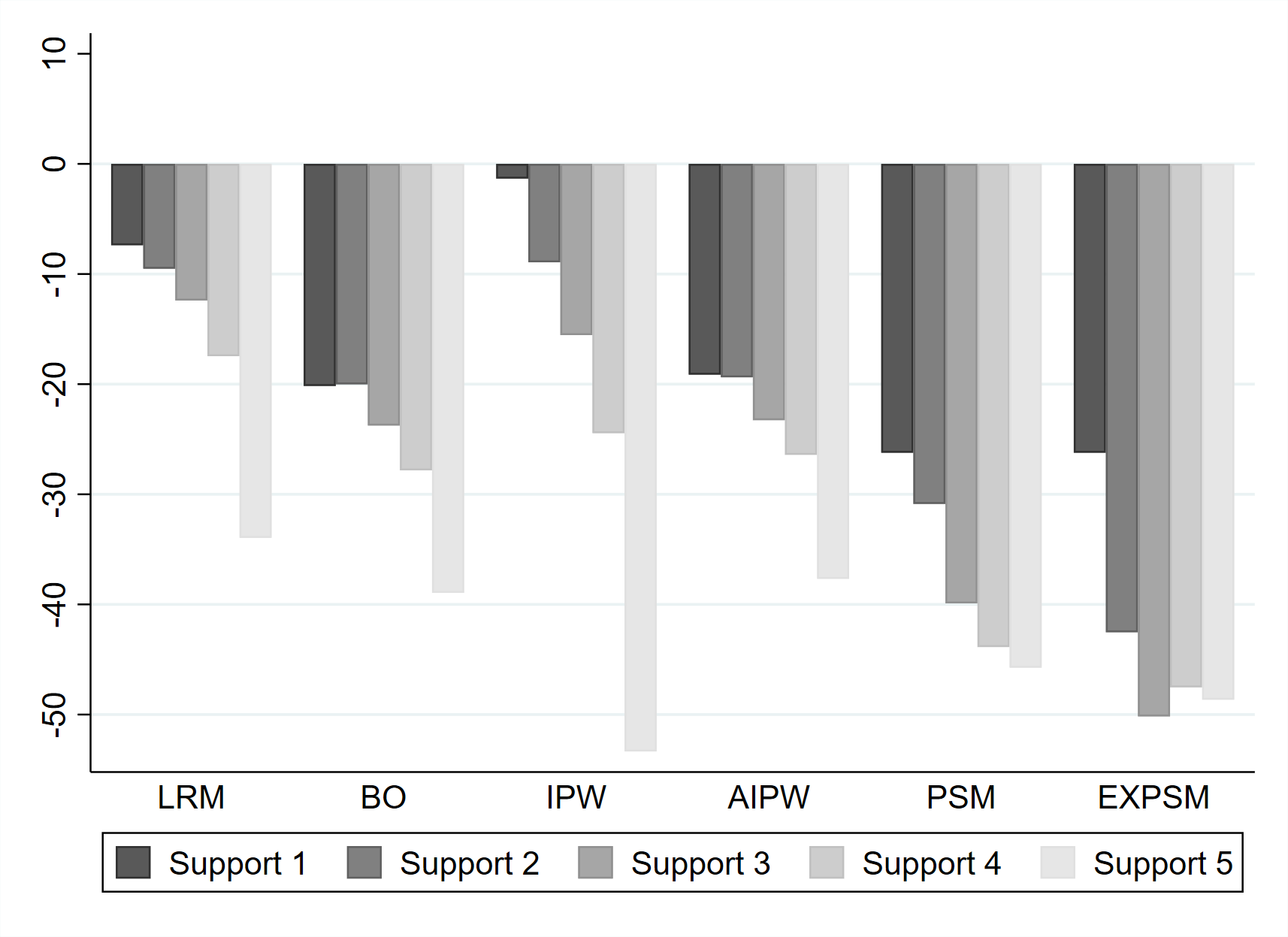}
\caption{Full model: difference to benchmark}
\end{subfigure}
\begin{subfigure}{0.45\textwidth}
\includegraphics[width=\textwidth]{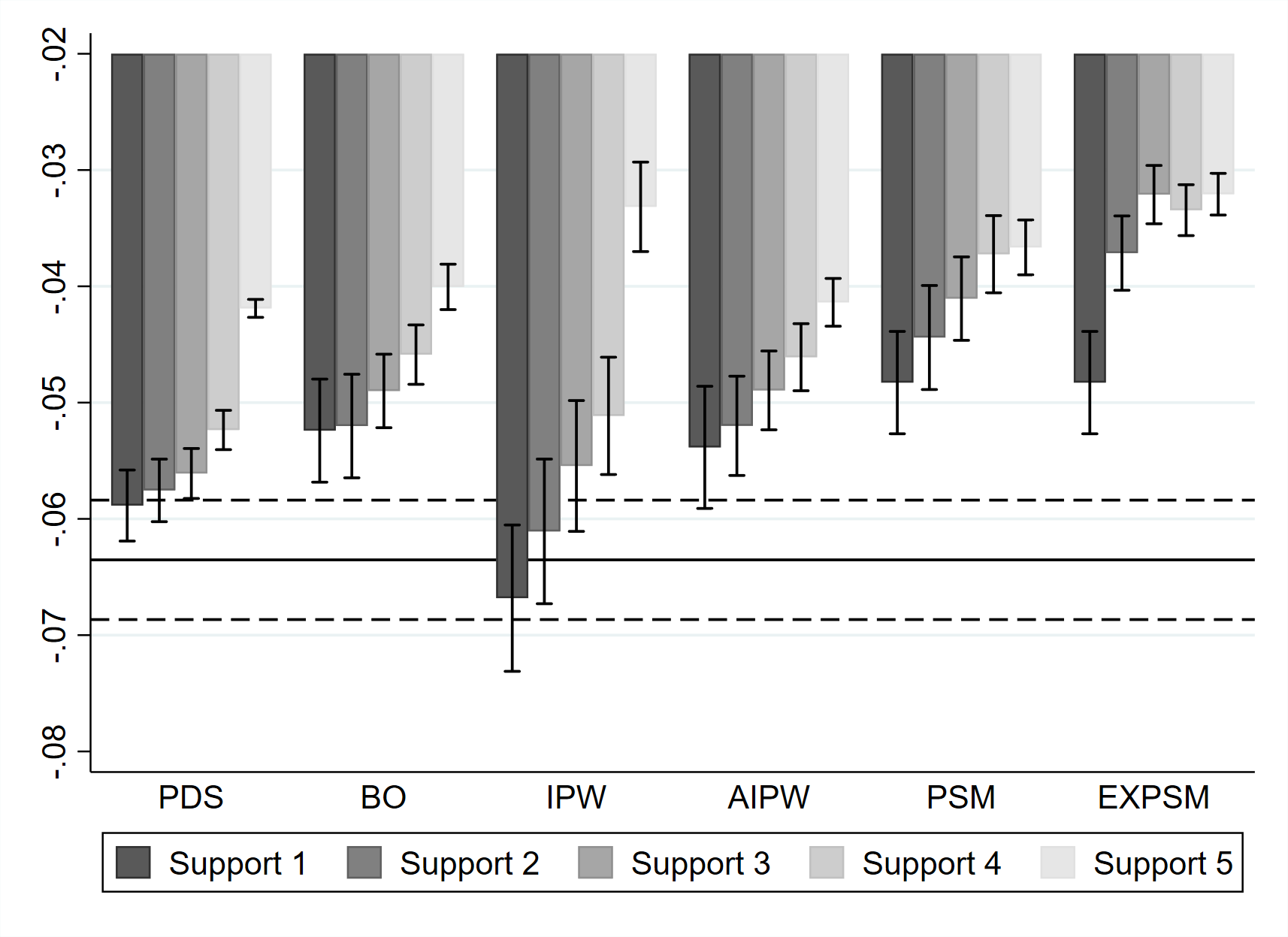}
\caption{ML model: estimates}
\end{subfigure}
\begin{subfigure}{0.45\textwidth}
\includegraphics[width=\textwidth]{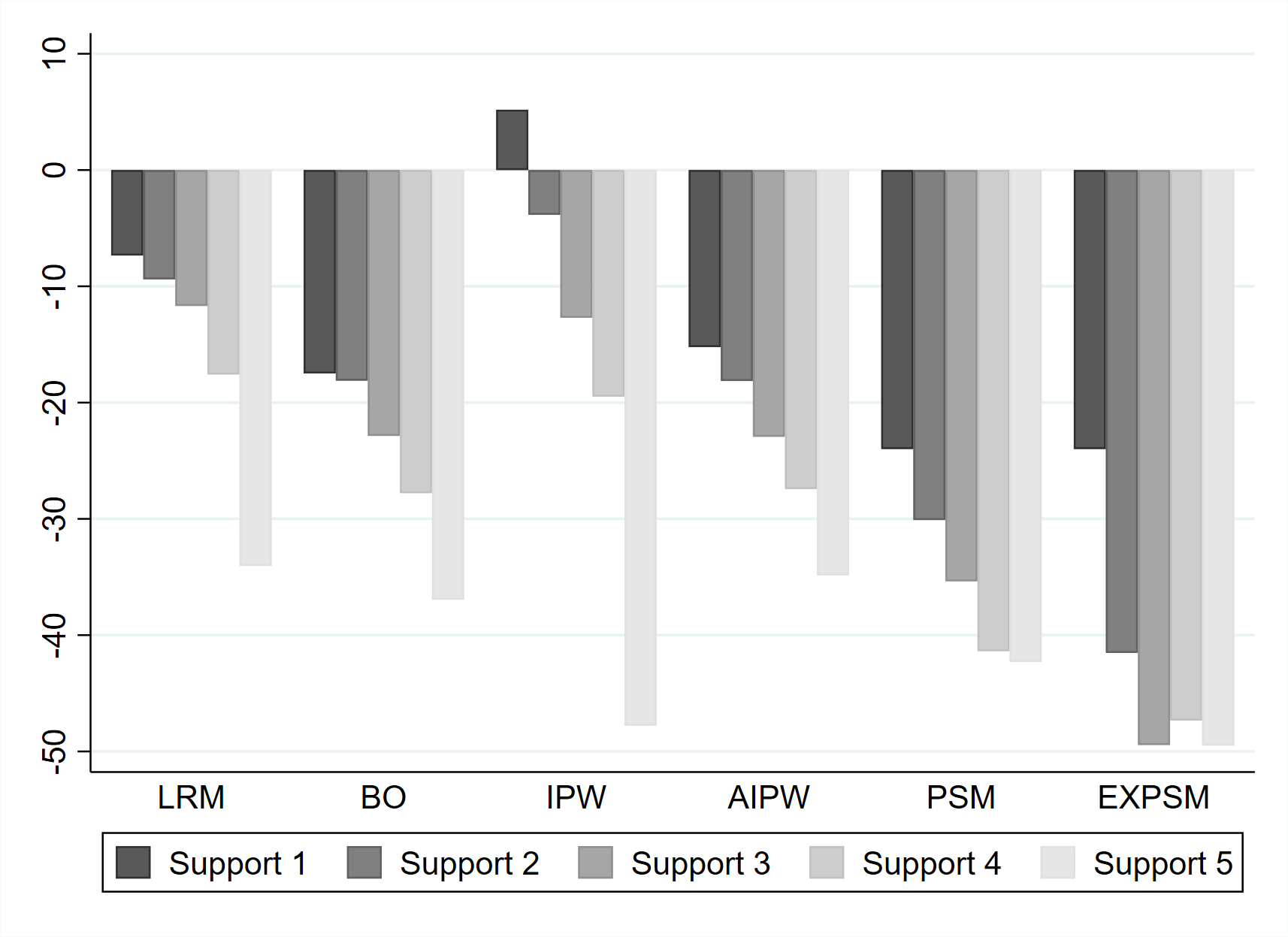}
\caption{ML model: difference to benchmark}
\end{subfigure}
\end{center}
{\footnotesize Notes:  We report all estimates and their standard errors in Table D.2 in Online Appendix D. Capped lines indicate the 95\% confidence intervals of the estimated unexplained gender pay gap. The black horizontal line in the left panels shows the BO estimate of Support 1 as a benchmark with the baseline specification and its 95\% confidence band in black dashed lines. The right panels show the difference from this benchmark in percent, i.e., the difference from the BO estimate using Support 1 with baseline model specification. PDS is the abbreviation for post-double-selection procedure.}
\end{figure}

{In panels (a), (c) and (e) of Figures \ref{results_graph_priv} and \ref{results_graph_pub}, we show the estimates of the average unexplained gender pay gap for the private and public sector, respectively. The bars show the unexplained gender pay gap estimates of the different methodological choices we consider and the capped lines show the 95\% confidence interval for each estimate.}\footnote{We do not include exact matching (EXM) as EXM does not control for all wage determinants in Supports (1)-(4), but EXM has already been discussed in the context of Figure \ref{cs_graph}. Tables D.1 and D.2 in Online Appendix D report all estimates including EXM with their standard error for the private and public sector, respectively.} As a benchmark, the black horizontal line shows the BO estimate in Support 1 (full sample except for two women in the public sector) with the baseline specification together with its 95\% confidence interval (black dashed lines). %This benchmark corresponds to what the majority of studies estimate. 
Furthermore, panels (b), (d) and (f) {of both figures} show the percentage differences of each estimate from this benchmark. 

We find that all of the methodological choices we consider matter greatly. Compared to the BO benchmark, using a less flexible estimator like the LRM increases the estimated unexplained gender pay gap by up to 29\%. In contrast, increasing flexibility by including wage determinants more flexibly or by using a more flexible estimator together with enforcing common support reduces the estimated pay gap by up to 50\%. These differences are substantial. In the following section, we investigate the role of each methodological choices separately: enforcement of common support, flexibility in the inclusion of wage determinants, and estimator choice.

\subsubsection{Role of stricter support enforcement} 
\begin{figure}[h!]
\begin{center}
\caption{Difference in estimated unexplained gender pay gap relative to Support 1} \label{diffsup}
\includegraphics[width=8.3cm]{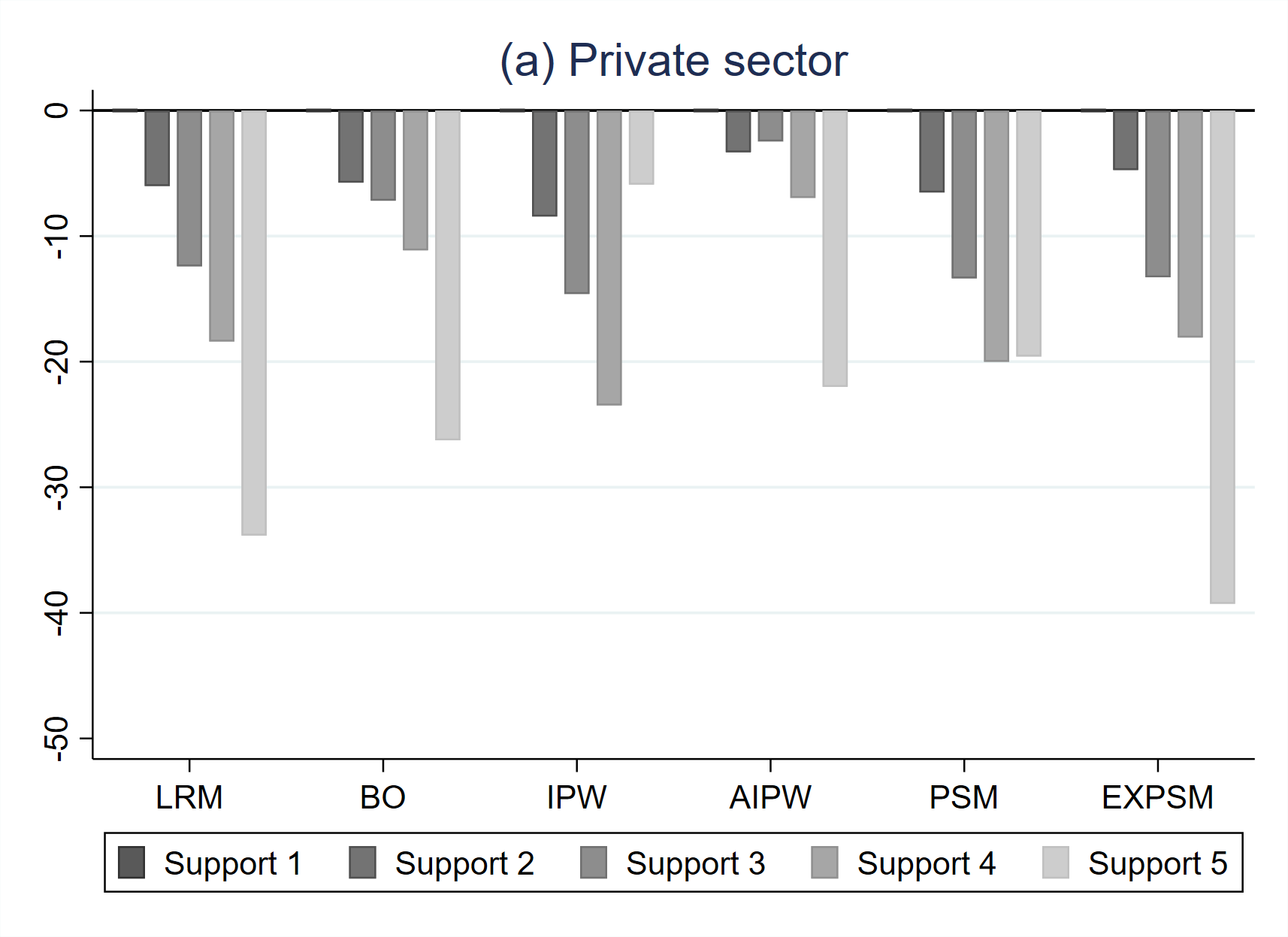}
\includegraphics[width=8.3cm]{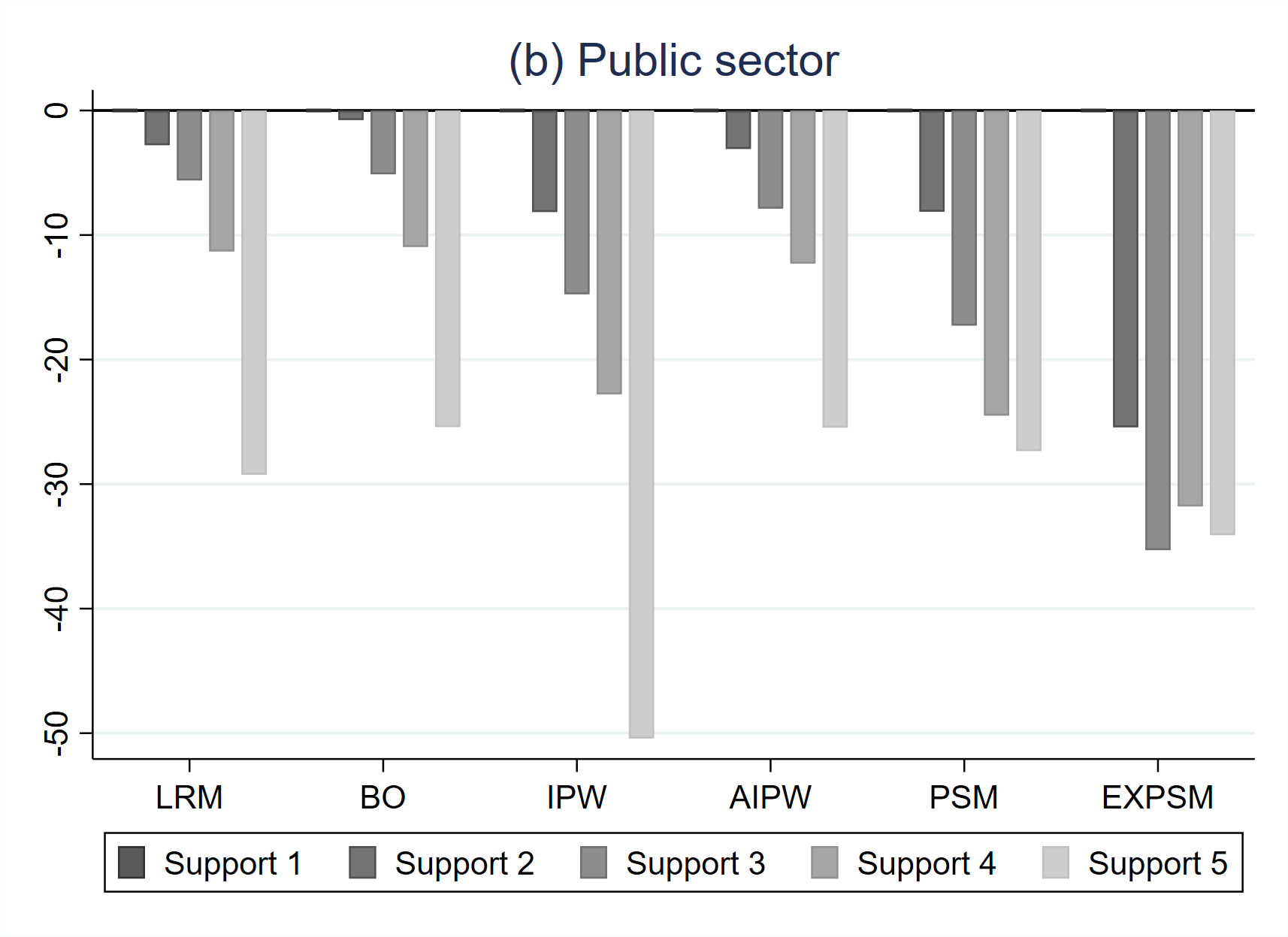}
\end{center}
{\footnotesize Notes: Difference in percent to the estimate for Support 1 using the respective estimator. The results for the model specifications (baseline, full, ML) are pooled.}
\end{figure}

Figure \ref{diffsup} shows the differences in the estimated unexplained gender pay gaps for each support definition relative to Support 1. We report separate results for each estimator, but pool the results for the different model specifications (baseline, full, ML) for a better overview. 

The estimated pay gaps shrink substantially and, with few unsystematic exceptions, monotonically with stricter support enforcement. For the BO estimates, enforcing support with respect to all quantitatively important wage determinants (Support 3) reduces the estimated pay gap by around 6\% in both sectors, while enforcing full support with respect to all variables (Support 5) reduces estimated gaps by around 26\%. With 13-35\% for Support 3 and 5-50\% for Support 5, the differences are considerably larger for the semi-parametric estimators IPW, PSM and EXPSM. Overall, enforcing common support has a very strong impact on the size of the estimated unexplained gender pay gaps.

There are two possible explanations for this finding. First, stricter support enforcement makes women and men more comparable. Hence, differences in observed wage determinants are likely to explain a larger share of the raw gender pay gap. This is particularly true for the public sector, where a large part of the raw gender pay gap can be explained by lack of support directly (see Figure \ref{cs_graph}). Second, heterogeneity in unexplained gender pay gaps can explain the differences across support samples that differ increasingly in their composition (see Table B.6 in Online Appendix B). For example, stricter support enforcement increasingly removes women in lower-paying jobs. As a result, the unexplained gender pay gap could either increase or decrease, depending on whether the unexplained gender pay gap is larger for women in higher- or lower-paying jobs. However, the direction cannot be inferred from data, because we cannot estimate the unexplained gender pay gap for the omitted women who lack comparable men.

The results are consistent with the findings of \cite{nopo08}, who shows that 11\% of Peru's gender pay gap estimates can be explained by lack of common support with regard to age, education, marital status, and migration condition. We document that support violations can be much more severe when accounting for a richer set of wage determinants. Furthermore, we show that all estimators are affected by support violations, including the parametric estimators that rely on extrapolation. 

\subsubsection{Role of model flexibility} 

\begin{figure}[h!]
\begin{center}
\caption{Difference in estimated unexplained gender pay gap relative to baseline model} \label{diffmodel}
\includegraphics[width=8.3cm]{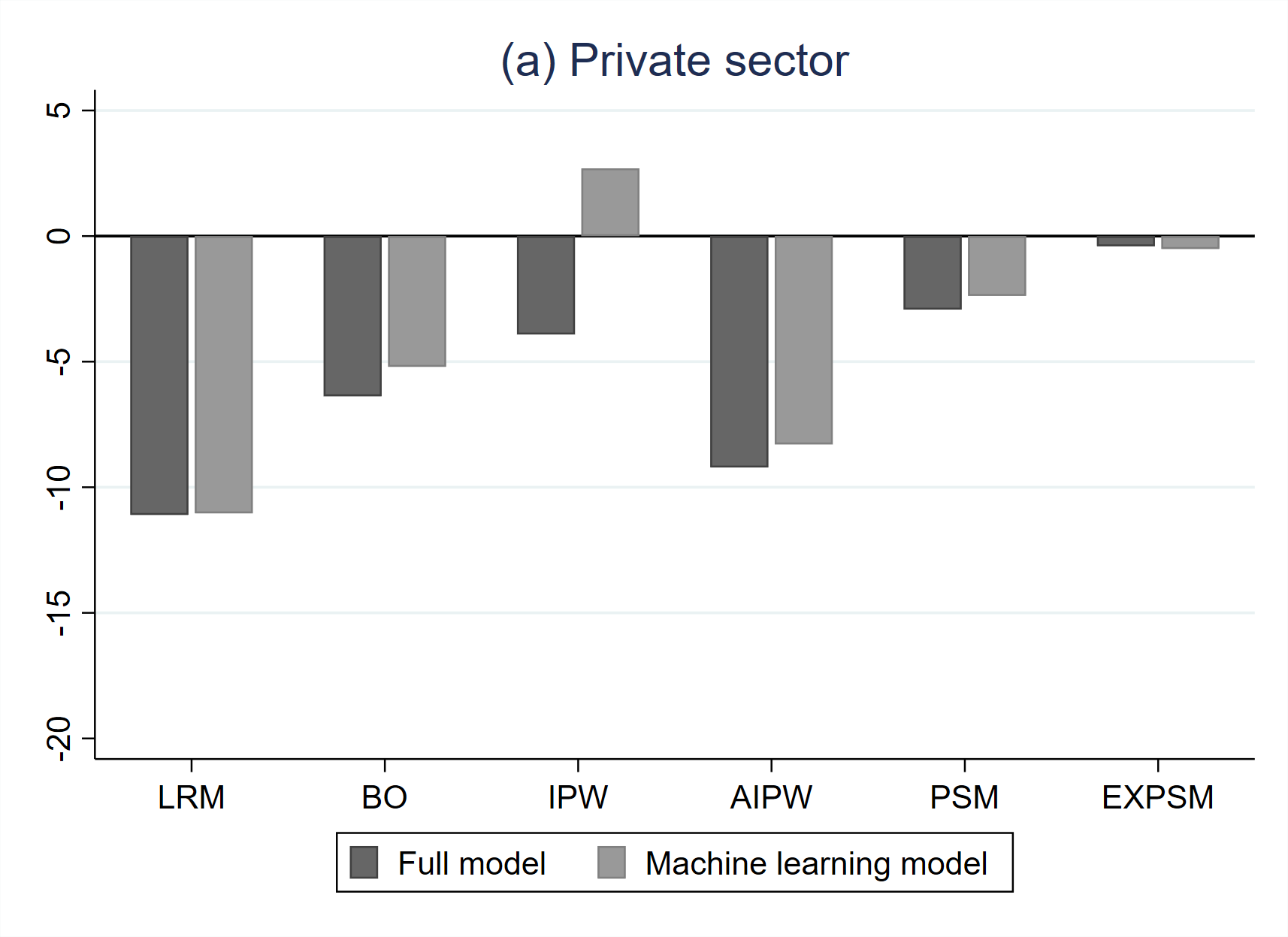}
\includegraphics[width=8.3cm]{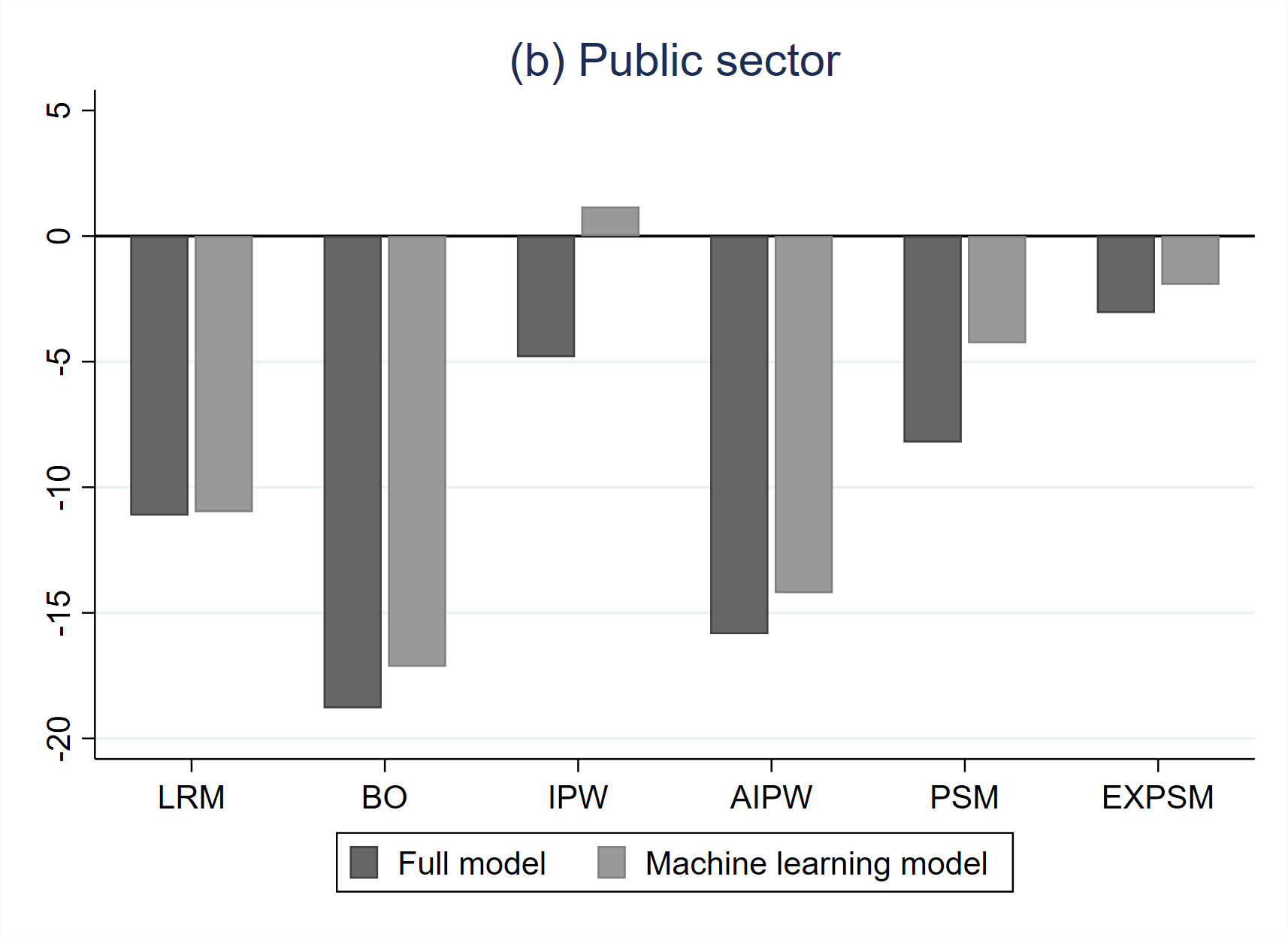}
\end{center}
{\footnotesize Notes: Difference in percent of the baseline model estimate with the respective estimator. The results for the different support definitions are pooled.}
\end{figure}

Controlling for wage determinants more flexibly is also important, especially for the parametric estimators. Figure \ref{diffmodel} shows the percentage differences in the estimated unexplained gender pay gaps between the baseline model of the respective estimator and the full and ML model specifications. We report separate results for each estimator, but pool the results of the different support definitions for a better overview. 

We find that the most flexible full model specification reduces the BO estimate of the pay gap by 6\% in the private sector and 19\% in the public sector. The reductions are only slightly smaller for the ML specification, which selects variables from the full model in a data-driven way to increase efficiency. 

Model flexibility matters most for the parametric estimators LRM and BO; as well as for AIPW, which can be viewed as an extension of BO with semi-parametric small-sample bias adjustment. All three estimators model the wage equation to estimate the unexplained gender pay gap. In contrast, the semi-parametric estimators IPW, PSM and EXPSM are much less sensitive to how wage determinants are included. All three estimators only use the propensity score as an input factor and do not impose any functional form on the wage equation. For those estimators, the estimated pay gaps differ by less than 8\% from the baseline model. The most flexible one, EXPSM, exhibits only very small differences across model specifications: only 0.4\% in the private sector and 3\% in the public. For all estimators, the difference between the full and ML models are not very large. A noticeable exception is the IPW estimator, where the unexplained gender pay gap of the ML specification exceeds even the baseline specification. \cite{kna20} document low finite sample performance of IPW in combination with ML, which may explain this result.

\cite{kline11} provides a theoretical justification for the sensitivity of the BO estimator with regard to model flexibility. He argues that BO is more vulnerable to model misspecification than IPW, because the implicit weighting scheme of BO allows for negative weights (which is not permitted for IPW). When the BO model is specified more flexibly, though, negative implicit weights become less likely.

We draw three conclusions from these results. First, controlling for wage determinants in a flexible way is important. Second, the more flexibly they are included, the better; misspecification of functional forms is less likely and the loss of degrees of freedom is not costly in our very large data set (but this might be different in smaller data sets). Third, flexible inclusion of wage determinants is more important for the estimators that incorporate the wage equation than for estimators that only incorporate the propensity score. With our very large database, we also find that applying machine learning methods for variable selection from a very rich set of non-linear and interaction terms has only a small impact on the estimated pay gaps. This may be different, though, in smaller samples where efficiency is more of a concern.

\subsubsection{Role of estimator choice} 
\begin{figure}[h]
\begin{center}
\caption{Difference in estimated unexplained gender pay gap relative to BO in respective support} \label{diffest}
\includegraphics[width=8.3cm]{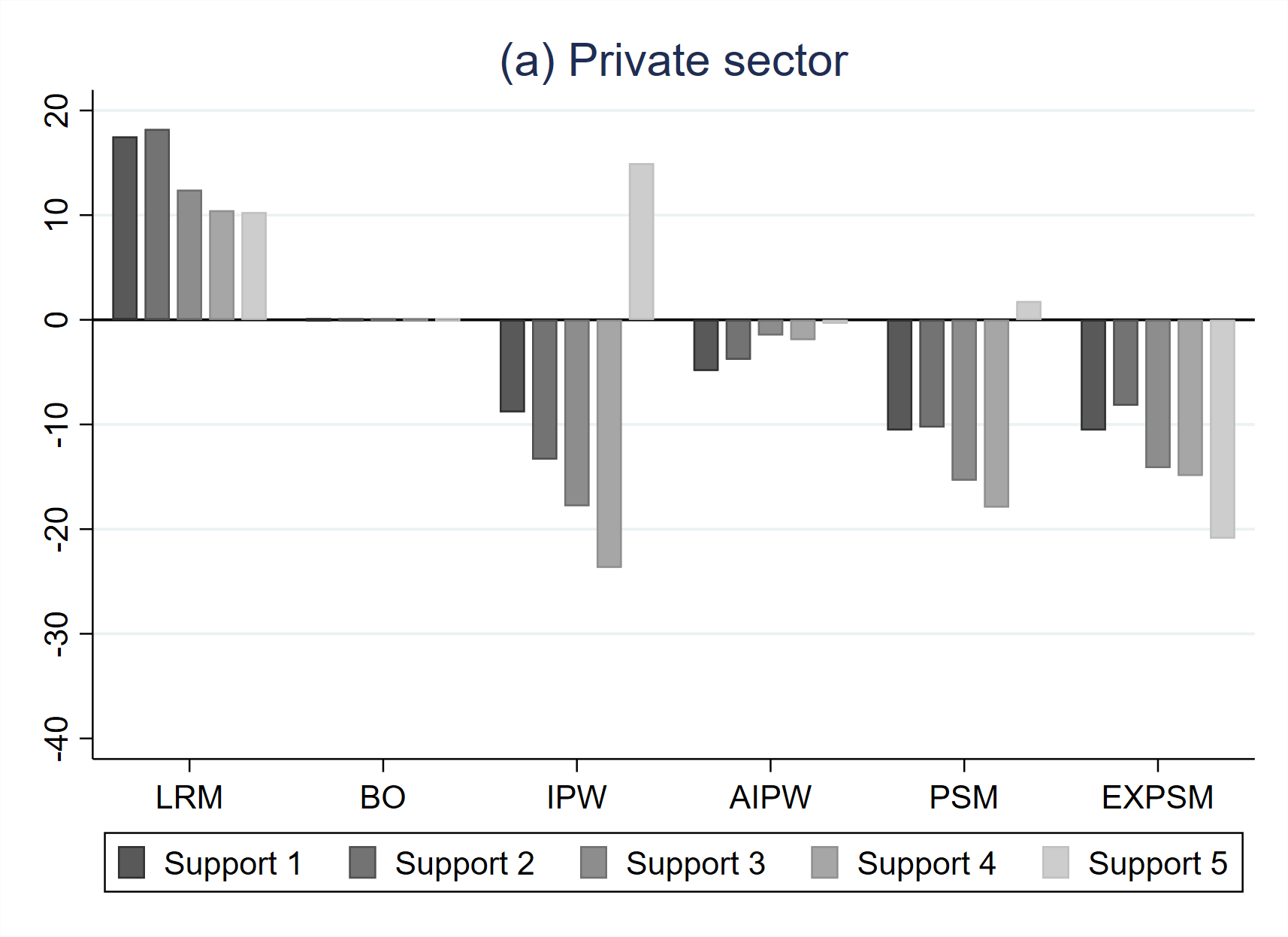}
\includegraphics[width=8.3cm]{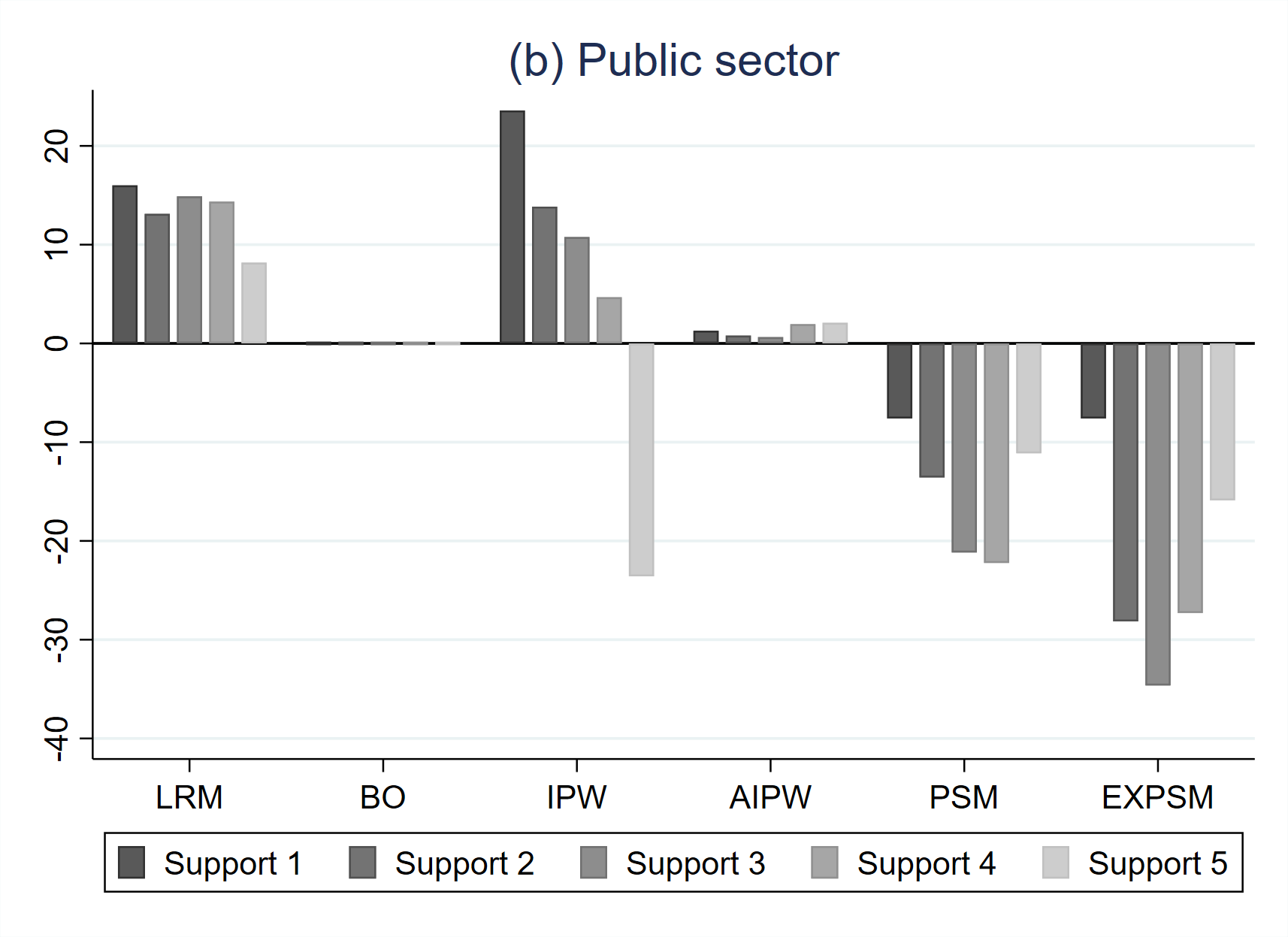}
\end{center}
{\footnotesize Notes: Difference in percent of the BO estimate in the respective support sample. All estimates are for the full model, which is the most flexible specification.}
\end{figure}

The last dimension we vary is estimator choice. Figure \ref{diffest} shows the differences in the estimated unexplained gender pay gaps between BO and the respective estimator for each support. As the flexibility of including wage determinants affects the parametric and semi-parametric estimators differently, we focus on the results from the full model with maximum flexibility for all estimators. This implies that the differences that we observe across estimators mainly result from differences in the way the estimators restrict possible heterogeneity in unexplained gender pay gaps. The LRM imposes homogeneous gender pay gaps. BO allows for heterogeneous gender pay gaps that are driven by gender differences in the returns to the included wage determinants. In contrast, the semi-parametric estimators IPW, PSM and EXPSM do not restrict heterogeneity in the pay gaps at all. AIPW is a mixture between the parametric BO and a small sample bias adjustment based on the the semi-parametric IPW. With our very large samples, small-sample bias should be negligible and BO and AIPW should yield very similar results.

We find that for each support, LRM estimates of pay gaps are 8-18\% higher than the BO estimate. The differences between the estimated BO and AIPW pay gaps are small as expected with our large samples. This may be different in smaller samples, though. For the matching estimators PSM and EXPSM we estimate pay gaps that are systematically and substantially smaller than the corresponding BO estimates on the same support by up to 20\% in the private and up to 34\% in the public sector. This is in line with substantial heterogeneity in estimated pay gaps that is widely acknowledged in the literature \citep[e.g.][]{bach18,chern18,goldin14}. 

However, we find no systematic pattern for IPW. In the private sector, we find estimates smaller than with BO and similar to PSM and EXPSM for supports 1-4, but much larger estimates for support 5. In the public sector, we obtain estimates that are 5-24\% larger than the BO estimates for supports 1-4 and much smaller for full support 5. The results for IPW show that this estimator is very sensitive to the population studied.

We conclude that it is important not to restrict pay gap heterogeneity by using a semi-parametric estimator. Moreover, among the semi-parametric estimators, PSM and EXPSM are much more robust and precise. The differences between PSM and EXPSM are small in most cases. With a sufficient numbers of observations, EXPSM is preferable because it ensures comparability of women and men in the best way among all considered estimators. Moreover, EXPSM is the least sensitive to the way we include the observed wage determinants because exact matching on the variables that define support fully removes all differences in these variables without imposing any restrictions.

These findings are consistent with the results of \cite{f07}, who documents that the PSM estimates of the unexplained gender pay gap in the UK
are up to 29\% smaller than the BO estimates. \cite{black08} also find strong differences between BO and EXM estimates of the unexplained gender pay gap in the United States. In particular, the estimates decline by 18\% for white women, 92\% for Hispanic women, and 83\% for Asian women. However, for black women they find no strong difference between the BO and EXM estimates. Likewise, \cite{nopo08} and \cite{gtv17} find no strong differences in the BO and EXM estimates of the unexplained gender pay gap in Peru and Poland, respectively, but they only account for a very small set of wage determinants. In Figure \ref{cs_graph}, we show that the selection of control variables is crucial for EXM.

\section{Discussion}\label{sec:disc}
\subsection{Relevance of the Swiss case}
A legitimate question is whether the results we obtain with the Swiss data are relevant for other applications. In terms of data, ours are very similar to those of the European Union Structure of Earnings Survey (SES). The SES provides harmonized data on earnings in EU member states, candidate countries, and EFTA countries. It includes all of the wage determinants that have been used to define Supports 1-4 (but not 5), which covers all quantitatively important wage determinants. Moreover, like the Swiss data, actual work experience is not included. Studies for the US typically use survey data collected from individuals such as the Current Population Survey (CPS), the US Census, or the American Community Survey (ACS). Such surveys contain much richer information on individuals but potentially suffer from measurement error in self-reported wages. However, the key wage determinants we observe are included as well, and most studies use a similar set of variables. In terms of sample size, all of the data sets are also very large, containing several hundred thousand observations.

With respect to labor market institutions, Switzerland is a particularly interesting case. On the one hand, it has generous social insurance systems like many other European countries. On the other hand, it has a very flexible labor market that is much less regulated than that of other European countries, which makes it more comparable to countries like the US. Thus, Switzerland shares important features of both typically European labor markets and flexible US-type labor markets. The differences we find between the private and the public sectors in Switzerland are also likely to be relevant for other countries. Public sectors in most countries share the features that explain the differences in the Swiss case. They are are typically more homogeneous in terms of covered occupations with strong concentration in certain service sectors, they typically attract a higher share of women than the private sector, they exhibit higher shares of high-skilled workers, and they are typically more regulated, including their wage setting, than private sectors \citep[see][]{brindusa2012}.

In summary, the data and labor market institutions in Switzerland are in many ways comparable to those in other countries. Therefore, we expect that our qualitative results extend to other settings as well. Of course, the magnitudes of gender pay gaps differ greatly across countries \citep[see e.g.][]{ww05,vts15}. Hence, we also expect the quantitative impact of methodological choices to differ. However, given that most countries share key features of female employment and gender segregation in the labor market, we are confident that the methodological choices we study are relevant beyond the Swiss context.

\subsection{Implications for applied research}

Our results suggest that deciding how to enforce common support is the first important choice in applied research about the gender pay gap. Researchers face a trade-off between stricter support to improve ex-ante comparability, and sample size. The latter affects both the precision of the estimate and the representativeness of the study sample. We recommend starting with an analysis of common support using the approach of \citet{nopo08}, adding variables in decreasing order of their impact on counterfactual male wages. Based on this analysis, there are two possible criteria for choosing the set of variables for enforcing support. One option is to use all quantitatively important wage determinants. In our application, this would result in support definition 3 with 39\% of employed women lost in the private sector and 20\% lost in the public sector, which is quite substantial. Alternatively, one could base the decision on the importance of wage determinants for explaining the raw gender pay gap rather than the counterfactual male wage. In this case, the researcher could choose the point at which the unexplained gender pay gap obtained from exact matching stabilizes. In our data, this is the case somewhere between Supports 2 and 3 (see Figure \ref{cs_graph}). As sample size drops considerably after enforcement of Support 2 plus irregular payments, this would be a natural choice of the set of variables for enforcing support. This would exclude 19\% of employed women in the private sector and 10\% in the public sector with the composition of the study sample being quite similar to that of the full sample. 

The next important decision is which estimator to use. The results indicate that restricting how observed wage determinants and gender may affect wages can have a large impact on the estimated unexplained gender pay gap. When using BO, including wage determinants in a flexible way is crucial. However, with a sufficiently large sample, a flexible matching estimator is even better, as matching does not restrict gender pay gap heterogeneity in any way. The results suggest that combining exact matching on few very important wage determinants that also define support with radius matching on a flexibly specified propensity score works particularly well. It minimizes the risk of functional form misspecification and, with support defined as described in the last paragraph, offers a good balance between ensuring comparability, precision and representativeness of the study sample.

Implementing these recommendations with our data would affect the estimated unexplained gender pay gap in the following way. Consider Support 3 as the more conservative of the two alternatives discussed above. In the private sector, enforcing Support 3 reduces the raw gender pay gap of 18.6\% by only 1\%. Standard BO with the baseline specification for including wage determinants that ignores support explains 58\% of the raw gap in the full sample and results in an estimate of the unexplained pay gap of 7.7\%. Enforcing Support 3 reduces this estimate by 5\%, using the most flexible instead of the baseline specification for BO by another 5\%, and using semi-parametric EXPSM with the same flexible specification instead by another 13\%. In total, the methodological choices decline the unexplained gender pay gap estimates by 23\%. Flexible EXPSM yields an estimated unexplained gap of 6\% and explains 68\% of the raw wage.

The differences are even larger for the public sector. Enforcing Support 3 already reduces the raw gender pay gap of 13.9\% by 14\%. Standard BO without support enforcement explains 54\% of the raw gap in the full sample and yields an estimate of the unexplained pay gap of 6.4\%. Enforcing Support 3 reduces this estimate by 4\%. Using the most flexible instead of the baseline specification for BO has a strong impact, as it reduces the estimate by another 19\%. Using semi-parametric EXPSM with the same flexible specification instead leads to another even more substantial reduction of 26\%. In total, the methodological choices decline the estimates of the unexplained gender pay gap by 50\%. Flexible EXPSM yields an estimated unexplained gap of only 3.2\% and explains 77\% of the raw wage. This once again illustrates the importance of these methodological choices.

\section{Conclusion}\label{sec:con}

We study the sensitivity of estimates of the unexplained gender pay gap to three types of methodological choices: enforcement of comparability of men and women ex ante, flexibility in the inclusion of wage determinants, and choice of estimator. We find that all of these choices matter greatly. Implementing the choices we recommended based on our results using data for the Swiss private sector, explains 16\% more of the raw wage gap than standard BO estimates and results in estimated unexplained pay gaps that are 23\%  lower. For the public sector, the preferred set of choices explains 42\% more of the raw wage gap than standard BO estimates and results in estimated unexplained pay gaps that are 50\% lower.

An important takeaway from our study is that the commonly reported Blinder-Oaxaca estimates of the unexplained gender pay gap can be misleading for several reasons. There is a high risk that they compare women to non-comparable men to an extent that is quantitatively  important. Moreover, they hide gender inequalities that result from this lack of comparability. For females with lack of support, we know nothing about possible inequalities in pay. Moreover, lack of support shows the extent of labor market segregation as an additional source of gender inequality. This is ignored if researchers do not check for common support. More generally, the restrictions standard BO imposes bear a high risk of yielding biased estimates of the unexplained gender pay gap. In particular, they underestimate the full extent of heterogeneity in pay gaps that is important for targeting affirmative action campaigns. Future work should study this heterogeneity in more detail.

As a final remark, we would like to emphasize that our findings provide guidance on how to estimate unexplained gender pay gaps with a given set of observed wage determinants. The results from such estimations are informative about equal pay for equal work taking individual choices as a given, subject to any omitted variable bias that may result from unobserved factors. Actual experience, having children, and personality traits are examples of such unobserved factors in the data we have used. Uncovering the extent and sources of possible gender discrimination in the labor market requires much more than accounting for unobserved wage determinants, though, such as dealing with selection into employment and endogenous control variables. 
\pagebreak

\bibliographystyle{AER}
\bibliography{bibliothek}

\end{document}